\documentclass[preprint]{aastex}
\usepackage{amsmath}
\usepackage{graphicx}
\usepackage{booktabs,threeparttable}
\usepackage{hyperref}
\usepackage{multirow}
\usepackage{ulem}
\usepackage{lscape}
\usepackage{textpos}
\shorttitle{Formation of Magnetized Prestellar Cores}
\shortauthors{Chen \& Ostriker}
\begin{document}
\title{Anisotropic Formation of Magnetized Cores in Turbulent Clouds}
\author{Che-Yu Chen\altaffilmark{1} and Eve C. Ostriker\altaffilmark{1,2}}
\altaffiltext{1}{Department of Astronomy, University of Maryland, College Park, MD 20742}
\altaffiltext{2}{Department of Astrophysical Sciences, Princeton University, Princeton, NJ, 08544}
\email{cychen@astro.umd.edu, eco@astro.princeton.edu}

\begin{abstract}

In giant molecular clouds (GMCs), shocks driven by converging turbulent flows create high-density, strongly-magnetized regions that are locally sheetlike.
In previous work, we showed that within these layers, dense filaments and embedded self-gravitating cores form by gathering material along the magnetic field lines. Here, we extend the parameter space of our three-dimensional, turbulent MHD core formation simulations. We confirm the anisotropic core formation model we previously proposed, and quantify the dependence of median core properties on the pre-shock inflow velocity and upstream magnetic field strength.
Our results suggest that bound core properties are set by the total dynamic pressure (dominated by large-scale turbulence) and thermal sound speed $c_s$ in GMCs, independent of magnetic field strength. 
For models with Mach number between $5$ and $20$, the median core masses and radii are comparable to the critical Bonnor-Ebert mass and radius defined using the dynamic pressure for $P_\mathrm{ext}$. 
Our results correspond to $M_\mathrm{core} = 1.2{c_s}^4 (G^3 \rho_0{v_0}^2)^{-1/2}$ and $R_\mathrm{core} = 0.34{c_s}^2 (G \rho_0{v_0}^2)^{-1/2}$ for $\rho_0$ and $v_0$ the large-scale mean density and velocity. 
For our parameter range, the median $M_\mathrm{core}\sim 0.1-1 M_\odot$, but a very high pressure cloud could have lower characteristic core mass. 
We find cores and filaments form simultaneously, and filament column densities are a factor $\sim 2$ greater than the surrounding cloud when cores first collapse.
We also show that cores identified in our simulations have physical properties comparable to those observed in the Perseus cloud. Superthermal cores in our models are generally also magnetically supercritical, suggesting that the same may be true in observed clouds.

\end{abstract}

\section{Introduction}
\label{sec:intro}

Prestellar core formation in giant molecular clouds (GMCs) is an important issue in theoretical studies of star formation, because these cores are the immediate precursors of protostars \citep{1987ARA&A..25...23S,2007ARA&A..45..565M,2013PPVI...Andre}. It is believed that the magnetic field and supersonic turbulence may both play important roles in core formation and subsequent evolution. In GMCs, simulations starting more than a decade ago have shown that overdense structures generated by supersonic turbulence may collapse gravitationally to form protostellar systems, while also attracting material from their surroundings \citep[e.g.][]{1999ApJ...527..285B,1999ApJ...513..259O,2000ApJ...535..887K,2001ApJ...553..227P,2003MNRAS.339..577B}. Magnetic fields limit compression in large-scale turbulence-induced shocks, channel material toward forming filaments, provide support for cores as they grow, and remove angular momentum in collapsing cores \citep{1956MNRAS.116..503M,1966MNRAS.132..359S,1976ApJ...210..326M,1985prpl.conf..320M,1991ApJ...373..169M, 2003ApJ...599..363A, 2009NewA...14..483B,2010ApJ...720L..26L,2013PPVI...Li}.

Because GMCs are only lightly ionized, and magnetic fields are only coupled to charged particles, magnetic stresses are mediated by ion-neutral collisions, and are affected by the level of ambipolar diffusion. Analytic studies and numerical simulations have shown that supersonic motions accelerate ambipolar diffusion \citep{2002ApJ...570..210F,2004ApJ...603..165H,2004ApJ...609L..83L}. Similar simulations with both strong turbulence and ambipolar diffusion have also demonstrated core evolution times, efficiency of star formation, and core structure similar to observations \citep{2005ApJ...631..411N,2008ApJ...687..354N,2008ApJ...679L..97K,2011ApJ...728..123K,2009NewA...14..483B}. More recently, \cite{2012ApJ...744..124C} studied the one-dimensional C-type shocks and identified a transient stage of turbulence-accelerated ambipolar diffusion. This transient stage, with timescale $t_\mathrm{transient}\sim 0.1-1$~Myr (depending on ionization), can explain the enhanced diffusion rate and affect the magnetization of cores that form.

In \citet[][hereafter CO14]{2014ApJ...785...69C}, we applied three-dimensional numerical simulations to study the roles of magnetic fields and ambipolar diffusion during prestellar core formation in turbulent cloud environments. Our simulations adopted the model framework of \cite{2011ApJ...729..120G} to focus on the shocked layer produced by turbulent converging flows, and surveyed varying ionization and angle between the upstream flow and magnetic field. In simulations, we found hundreds of self-gravitating cores with masses $M\sim 0.04-2.5$~M$_\odot$ and sizes $L\sim 0.015-0.07$~pc, all formed within $1$~Myr. 

In \hyperlink{CO14}{CO14}, we also found that core masses and sizes do not depend on either the ionization or upstream magnetic field direction, and ambipolar diffusion is in fact not necessary to form low-mass supercritical cores. Our analysis showed that this is the result of anisotropic contraction along field lines, which can be clearly seen in our simulations, with or without ambipolar diffusion (see also \citealt{{2014ApJ...789...37V}}). 
In the anisotropic core formation model, low-mass magnetically supercritical cores form rapidly even in a strongly magnetized medium with high ionization. This explains the prevalence of magnetically supercritical cores in observations \citep{2012ARA&A..50...29C}.

Using a simple scaling argument, \hyperlink{CO14}{CO14} suggested the characteristic core mass may be set by the mean turbulent pressure in a GMC, regardless of magnetic effects. 
The predicted core mass is a factor $\sim {\cal M}^{-1}$ lower than the thermal Bonnor-Ebert mass at the mean density in the cloud, where ${\cal M}$ is the turbulent Mach number. Previously, \cite{2011ApJ...729..120G} proposed a similar formula based on the preferred scale for gravitational fragmentation of post-shock layers, for the purely hydrodynamic case. \cite{1997MNRAS.288..145P} also argued for a similar characteristic mass, based on statistics of turbulent flows. Although the analyses of \cite{2011ApJ...729..120G} and \cite{1997MNRAS.288..145P} neglect magnetic fields, the end result is similar to the prediction of \hyperlink{CO14}{CO14} that incorporates magnetic effects, with the turbulent pressure in a cloud setting the characteristic core mass.

Here, following \hyperlink{CO14}{CO14}, we continue our study of anisotropic core formation in turbulent molecular clouds. We extend our previous numerical study to explore how the turbulent and magnetic pressures of the pre-shock gas can affect core formation in the compressed region. We demonstrate that the dependence of core properties on pre-shock parameters are similar to those predicted by the anisotropic core formation model of \hyperlink{CO14}{CO14}. We also compare our results with observations, showing that the mass-size relationship and ratio of mass to critical value of our simulations is comparable to that seen in Perseus and other star-forming regions \citep{2010ApJ...710.1247S,2013MNRAS.432.1424K}.

The outline of this paper is as follows. We review the anisotropic core formation model in Section~\ref{sec::aniCore}, outlining the successive dynamical stages and associated parameter dependence expected. Section~\ref{sec::methods} describes the equations solved in our numerical simulations, and specifies the model parameter set we shall consider. The post-shock gas structure for varying parameters is analyzed in Section~\ref{sec::postshock}, including physical properties of the compressed layer (Section~\ref{sec::pslayer}), and development of filaments within it (Section~\ref{sec::psFila}). In Section~\ref{sec::cores} we provide quantitative results for masses, sizes, magnetizations, and other physical properties of the bound cores identified from our simulations, and compare to predictions from \hyperlink{CO14}{CO14}. We also compare these results with observations (Section~\ref{sec::obs}), focusing on interpreting the physical state of super-Jeans mass cores and mass-size relationships.
Section~\ref{sec::summary} summarizes our conclusions.

\section{Anisotropic Core Formation: Review}
\label{sec::aniCore}

\begin{figure}[t]
\begin{center}
\includegraphics[scale=0.6]{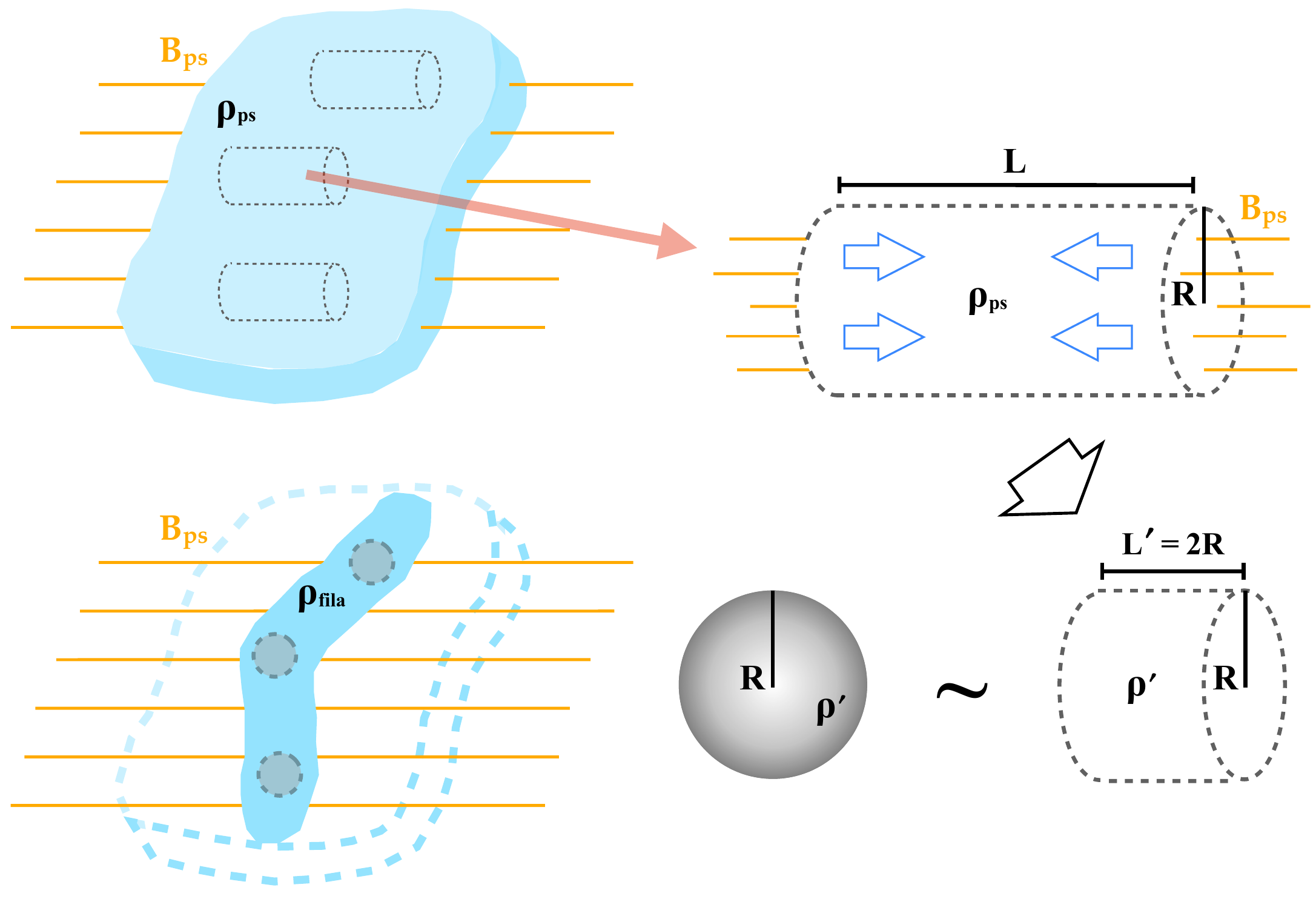}
\caption{Schematic of the anisotropic condensation process. The blue region {\it (top left)} shows a section of the post-shock layer created by a converging flow. Contraction initially begins along the direction of the post-shock magnetic field, which is nearly parallel to the post-shock layer. This contraction is indicated for a cylinder of initial length $L$ and radius $R$ {\it (top right)}. When the length of the cylinder has shrunk to satisfy $L' \sim 2R$, it can be treated as an isotropic sphere with radius $R$, which will collapse if the self-gravity overcomes thermal pressure. Contraction along the post-shock magnetic field creates dense filaments, and the densest regions within the filament continue contracting as quasi-spherical cores {\it (bottom left)}.}
\label{anisotropic}
\end{center}
\end{figure}

Here we briefly review the anisotropic condensation model of core formation proposed in \hyperlink{CO14}{CO14}. We consider a strongly-magnetized post-shock region created by a large-scale converging turbulent flow within a cloud.
As shown in \hyperlink{CO14}{CO14} (see Figures~3$-$5 there), the magnetic field will lie primarily parallel to the shock front in the layer no matter what the initial inclination angle is, because only the component of the field parallel to the shock front (or perpendicular to the inflow) is amplified. 
Equation~(4) of \hyperlink{CO14}{CO14} gives the compression ratio for the component of the magnetic field that is amplified.
The anisotropic condensation model describes core formation in the post-shock layer with density $\rho_\mathrm{ps}$ and threaded by magnetic field strength $B_\mathrm{ps}$ (Figure~\ref{anisotropic}, top left).
For a cylinder with radius $R$ and length $L$ along the magnetic field (Figure~\ref{anisotropic}, \textit{top right}), if $2R \lesssim L \lesssim L_\mathrm{mag,crit}$ (see Equations~(30) and (31) of \hyperlink{CO14}{CO14}) for
\begin{equation}
L_\mathrm{mag,crit} = \frac{B_\mathrm{ps}}{\rho_\mathrm{ps}}\frac{1}{2\pi\sqrt{G}},
\label{Lcrit}
\end{equation}
it is gravitationally stable to transverse contraction across the magnetic field \citep{1956MNRAS.116..503M}. However, the magnetic field does not prohibit contraction along the length of the cylinder, and gravity will be able to overcome pressure forces if $L$ exceeds the thermal Jeans length within the post-shock layer, $L_{J,\mathrm{2D}} \equiv {c_s}^2/ G\Sigma_\mathrm{ps}$ or $L_{J,\mathrm{3D}} \equiv c_s (\pi/G\rho_\mathrm{ps})^{1/2}$. Here, $\Sigma_\mathrm{ps}$ is the total surface density of the post-shock layer. 
In this situation, 
longitudinal contraction along a flux tube can continue until an approximately isotropic core with $L' \sim 2R$ is produced (Figure~\ref{anisotropic}, \textit{bottom}), with density
\begin{equation}
\rho' = \frac{L}{2R}\rho_\mathrm{ps}.
\label{newrho}
\end{equation}
At this point, transverse contraction is no longer impeded by the magnetic field provided the original $L\sim L_\mathrm{mag,crit}$
so the core is magnetically supercritical (note that the mass-to-magnetic flux ratio remains the same during the longitudinal contraction).
The core will also have sufficient gravity to overcome thermal pressure support provided its mass is comparable to that of a critical Bonnor-Ebert sphere at ambient density $\rho'$, which corresponds to radius (prior to central concentration)
\begin{equation}
R\sim R_\mathrm{th,sph} = 2.3\frac{c_s}{\sqrt{4\pi G\rho'}}.
\label{Raniso}
\end{equation}
Combining $L\sim L_\mathrm{mag,crit}$ with Equations~(\ref{Lcrit})-(\ref{Raniso}), this yields
\begin{equation}
\rho' = 0.19\frac{{B_\mathrm{ps}}^2}{4\pi{c_s}^2} \approx 0.38 \frac{\rho_0{v_0}^2}{{c_s}^2}.
\label{rhop}
\end{equation}

In Equation~(\ref{rhop}), we have assumed a strong magnetized isothermal shock with downstream magnetic pressure balanced by upstream ram pressure (${B_\mathrm{ps}}^2/\left(8\pi\right) \approx \rho_0 {v_0}^2$)\footnote{The post-shock magnetic, thermal, and dynamic pressures can be directly measured in the simulation. From values listed in Table~1, the magnetic pressure in the post-shock region (${B_\mathrm{ps}}^2/(8\pi)$) is comparable to the upstream ram pressure ($\rho_0 {v_0}^2$), while the thermal pressure ($\rho_\mathrm{ps} {c_s}^2$) and dynamic pressure ($\rho_\mathrm{ps} {v_\mathrm{rms}}^2$) in the post-shock layer are about $1-2$ orders of magnitude smaller. Therefore it is safe to say that the post-shock magnetic pressure is the dominant component that balances the upstream ram pressure. }
so that
\begin{equation}
B_\mathrm{ps} = 31.04~\mu\mathrm{G}\left(\frac{v_0}{1~\mathrm{km/s}}\right)\left(\frac{n_0}{10^3\mathrm{cm}^{-3}}\right)^{1/2},
\label{Bps}
\end{equation}
where $\rho_0$, $v_0$ are the density and inflow velocity of the shock, respectively, and $n_0 = \rho_0/\mu_n$ for $\mu_n = 2.3m_\mathrm{H}$ the mean molecular weight.
We can then solve for the critical radius and mass that allows an anisotropically formed core to be both magnetically and thermally supercritical:
\begin{align}
R_\mathrm{crit} &= 5.3\frac{{c_s}^2}{\sqrt{G}B_\mathrm{ps}} = 1.06\frac{{c_s}^2}{\sqrt{G\rho_0 {v_0}^2}}\notag \\
&= 0.09~\mathrm{pc} \left(\frac{n_0}{1000~\mathrm{cm}^{-3}}\right)^{-1/2}\left(\frac{v_0}{1~\mathrm{km/s}}\right)^{-1}\left(\frac{c_s}{0.2~\mathrm{km/s}}\right)^2,
\label{Rcrit}
\end{align}
and
\begin{align}
M_\mathrm{crit} & = 14\frac{{c_s}^4}{G^{3/2}B_\mathrm{ps}} = 2.8 \frac{{c_s}^4}{\sqrt{G^3 \rho_0 {v_0}^2}}\notag\\
& = 2.1~\mathrm{M}_\odot \left(\frac{n_0}{1000~\mathrm{cm}^{-3}}\right)^{-1/2}\left(\frac{v_0}{1~\mathrm{km/s}}\right)^{-1}\left(\frac{c_s}{0.2~\mathrm{km/s}}\right)^4.
\label{Mcrit}
\end{align}
Equation~(\ref{Mcrit}) uses $M_\mathrm{crit} = \pi {R_\mathrm{crit}}^2 L_\mathrm{mag,crit} \rho_\mathrm{ps} = {R_\mathrm{crit}}^2 B_\mathrm{ps}/(2\sqrt{G})$.
Equations~(\ref{Rcrit}) and (\ref{Mcrit}) suggest that the characteristic mass of prestellar cores formed in post-shock regions in magnetized GMCs is determined by the dynamical pressure in the cloud, independent of the cloud's magnetization, when anisotropic condensation along the magnetic field is taken into account.
\hyperlink{CO14}{CO14} already showed that models with varying upstream magnetic field directions have similar values of the median core mass and radius. Here, we extend our previous investigation to consider variation in the inflow velocities and background magnetic field strength.

\section{Numerical Methods and Models}
\label{sec::methods}

The simulation setup is similar to \hyperlink{CO14}{CO14}, and is summarized here. We employ a three-dimensional ideal MHD model with convergent flow, self-gravity, and a perturbed turbulent velocity field \citep{2011ApJ...729..120G}. These numerical simulations are conducted using the \textit{Athena} MHD code \citep{2008ApJS..178..137S} with the Roe Riemann solver. As we found in \hyperlink{CO14}{CO14} that ambipolar diffusion plays a secondary role in core formation, here we consider ideal MHD. The equations we solve are:
\begin{subequations}
\begin{align}
\frac{\partial\rho}{\partial t} &+ \mathbf{\nabla}\cdot\left(\rho\mathbf{v}\right) = 0,\\
\frac{\partial\rho\mathbf{v}}{\partial t} &+ \mathbf{\nabla}\cdot\left(\rho\mathbf{v}\mathbf{v} - \frac{\mathbf{B}\mathbf{B}}{4\pi}\right) + \mathbf{\nabla}P^* = 0,\\
\frac{\partial\mathbf{B}}{\partial t} &+ \mathbf{\nabla}\times\left(\mathbf{B}\times\mathbf{v}\right) = 0,
\end{align}
\end{subequations}
where $P^* = P + B^2/(8\pi)$. For simplicity, we adopt an isothermal equation of state $P=\rho {c_s}^2$ with $c_s=0.2$~km/s.
For both the whole simulation box initially and the inflowing gas subsequently, we apply perturbations following a Gaussian random distribution with a Fourier power spectrum $v^2\left(k\right)\propto k^{-4}$ (\citealt{2011ApJ...729..120G}). 
The amplitude of the velocity dispersion $\delta v$ thus depends on the Mach number of the inflow ${\cal M}$ (see Equations~(21) and (22) in \hyperlink{CO14}{CO14}), as $\delta v = 0.14~\mathrm{km/s}\cdot({\cal M}/10)^{1/2}$.

Our simulation box is $1$~pc on each side, representing a region within a GMC where a large-scale supersonic converging flow with velocity $\mathbf{v}_0 = v_0~\hat{\mathbf{z}}$ and $-\mathbf{v}_0$ (i.e. in the center-of-momentum frame) collides. The $z$-direction is the large-scale inflow direction, and we adopt periodic boundary conditions in the $x$- and $y$-directions. We initialize the background magnetic field in the cloud, $B_0$, in the $x$-$z$ plane, with an angle $\theta=20^\circ$ with respect to the convergent flow. The number density of the neutrals, defined as $n \equiv \rho/\mu_n$, is set to $n_0 = 1000$~cm$^{-3}$ in the initial conditions and in the upstream converging flow. The physical parameters defining each model are then the inflow Mach number and upstream magnetic field strength ${\cal M} \equiv v_0/c_s$ and $B_0$. We choose ${\cal M} = 5$, $10$, and $20$ to look at the dependence of core mass/size on the inflow velocity, and $B_0 = 5$, $10$, and $20~\mu$G to test whether the initial magnetization of the cloud can affect the core properties (see Table~\ref{simresults}).

Similar to our previous work, we repeat each model $6$ times with different random realizations of the same turbulent power spectrum to collect sufficient statistical information. Note that the resolution adopted in \hyperlink{CO14}{CO14} ($\Delta x = 1/256$~pc) is not high enough to resolve strong shocks generated by high inflow velocity, especially ${\cal M}=20$ cases. 
Therefore, we increased our resolution to $512^3$ for all models in this work, such that $\Delta x \approx 0.002$~pc. 
We use H-correction \citep{1998JoCoPh...145...511S} to suppress the carbuncle instability, and, when needed, first order flux correction \citep{2009ApJ...691.1092L}
(i.e.~dropping back to first-order flux estimates for updating the gas variables, if higher-order estimates results in negative density).

From each simulation, we apply the \textit{GRID} core-finding method \citep{2011ApJ...729..120G}, which uses the largest closed gravitational potential contours around single local minimums as core boundaries. We then select the gravitationally bound cores as those with negative total energy (sum of gravitational, magnetic, and thermal energy). It is then straightforward to measure the mass and size for each identified core. For the magnetic flux within a core, we first find the plane that includes the minimum of the core's gravitational potential and is perpendicular to the average magnetic field direction within the core. This plane separates the core into an upper half and a lower half, and we can measure the magnetic flux $\Phi_B$ through the core by summing up $\mathbf{B}\cdot \mathbf{\hat n}$ in either the upper or lower half of the core surface (see \hyperlink{CO14}{CO14}). The normalized mass-to-magnetic flux ratio of the core is therefore $\Gamma \equiv M/\Phi_B \cdot 2\pi\sqrt{G}$.

\renewcommand{\arraystretch}{1.1}
\begin{table}[t]
      \footnotesize
\hspace{-.3in}
  \begin{threeparttable}
\caption{Summary of the post-shock properties measured and derived from simulations.}
\label{simresults}
\vspace{.1in}
\begin{tabular}{ l || c c | c c c c | c c c c c }
  \hline
  \multirow{3}{*}{Model} & \multicolumn{2}{c|}{cloud conditions} & \multicolumn{4}{c|}{simulated post-shock properties\tablenotemark{\ast}} & \multicolumn{5}{c}{corresponding physical scales\tablenotemark{\dagger}} \\
  \cline{2-12}
  & $v_0$ & $B_0$ & $\overline{n}_\mathrm{ps}$ & $\overline{B}_\mathrm{ps}$ & $\beta_\mathrm{ps}$ & $v_\mathrm{rms}$ & $M_\mathrm{mag,sph}$ &  $R_\mathrm{mag,sph}$ & $M_\mathrm{th,sph}$& $R_\mathrm{th,sph}$ & $L_\mathrm{mag,crit}$ \\ 
   & (km/s) & ($\mu$G) & ($10^3$~cm$^{-3}$) & ($\mu$G) & & (km/s) & (pc) & (pc) & (pc) & (pc) & (pc) \\
  \hline 
  M5B10 & 1 & 10 & 12.1 & 43.0 & 0.25 & 0.44 & 7.64 & 0.14 & 1.26 & 0.07 &  0.19  \\
  M10B10  & 2 & 10 & 18.7 & 66.8 & 0.16 & 0.57 & 12.0 & 0.14 & 1.02 & 0.06  & 0.18  \\
  M20B10 & 4 & 10 & 32.6 & 121 & 0.09 & 0.79 & 23.5 & 0.14 & 0.77 & 0.05 & 0.19  \\
  \hline
  M10B5 & 2 & 5 &  31.8 & 59.3 & 0.35 & 0.41 & 2.90 & 0.07 & 0.78 & 0.05 & 0.10  \\
  M10B10  & 2 & 10 & 18.7 & 66.8 & 0.16 & 0.57 & 12.0 & 0.14 & 1.02 & 0.06 & 0.18  \\
  M10B20 & 2 & 20 & 9.89 & 70.9 & 0.08 & 0.98 & 51.3 & 0.28 & 1.40 & 0.08 & 0.37 \\
  \hline
\end{tabular}
    \begin{tablenotes}
      \item $^\ast$Post-shock properties are measured at $t=0.2$~Myr in each model, averaged over the whole post-shock layer. The timescale is chosen so the downstream properties are measured before the post-shock layer becomes strongly self-gravitating. 
      \item $^\dagger$See Equations~(\ref{Lcrit}) and (\ref{Rthsph})-(\ref{Mmagsph}) for definitions of physical scales.
      \end{tablenotes}
  \end{threeparttable}
\end{table}

\section{Post-shock Environment and Structure Formation}
\label{sec::postshock}
\subsection{Post-shock Layer}
\label{sec::pslayer}

\begin{figure}
\begin{center}
\includegraphics[width=\textwidth]{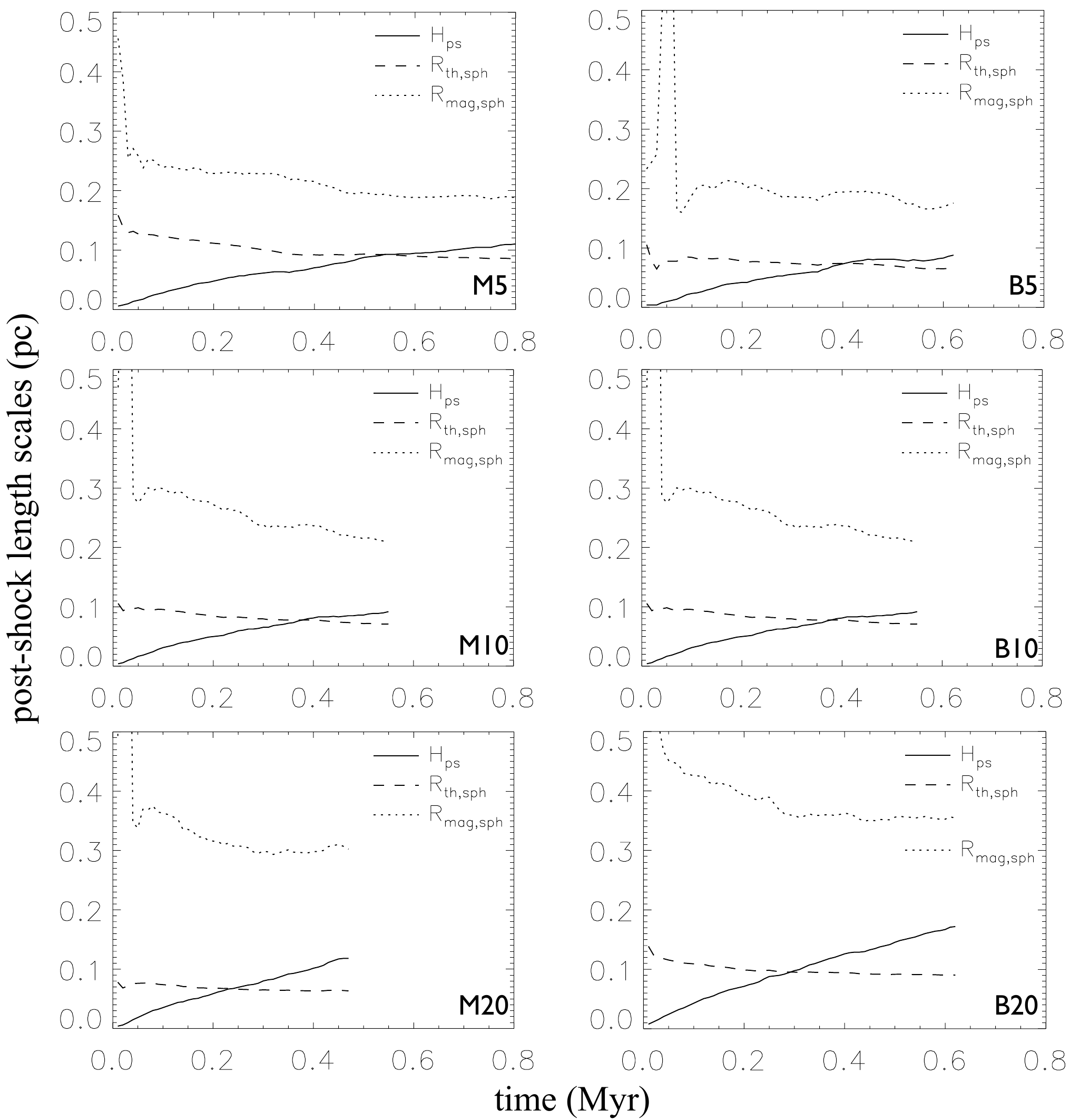}
\caption{The post-shock layer thickness $H_\mathrm{ps}$ in different models measured from simulations (\textit{solid}), compared to possible mass-gathering scales, $R_\mathrm{th,sph}$ from Equation~(\ref{Rthsph}) (\textit{dashed}) and $R_\mathrm{mag,sph}$ from Equation~(\ref{Rmagsph}) (\textit{dotted}) within the post-shock layer. Since the post-shock layer is strongly magnetized with $R_\mathrm{mag,sph}$ much larger than $H_\mathrm{ps}$ during the core building phase ($\sim 0.5$~Myr), cores cannot collect mass along the direction perpendicular to the layer. }
\label{HpsRsph}
\end{center}
\end{figure}

The post-shock results from our simulations are summarized in Table~\ref{simresults}. Similar to \hyperlink{CO14}{CO14}, we measured the post-shock properties at $t=0.2$~Myr, a timescale that is short enough that no cores have formed, yet long enough for the post-shock region to reach a steady-state solution\footnote{Although the physical conditions within the post-shock layer do not vary much in time during the initial stages of evolution, its thickness grows (approximately linearly in time) because mass is accumulated from the continual inflow. The evolution of the layer's thickness is consistent with the expectation $H_\mathrm{ps} = (2 \rho_0 v_0 / \rho_\mathrm{ps})t$ (see Figure~\ref{HpsRsph}).}
as derived in Section~2.1 of \hyperlink{CO14}{CO14}.
In fact, the timescale $t_\mathrm{sg}$ necessary for the post-shock layer to become self-gravitating can be derived by considering when the gravitational weight,
\begin{equation}
\frac{\pi G {\Sigma_\mathrm{ps}}^2}{2} = \frac{\pi G \left(2\rho_0 v_0 t_\mathrm{sg}\right)^2}{2},
\end{equation}
exceeds the post-shock pressure ${B_\mathrm{ps}}^2/8\pi \approx \rho_0 {v_0}^2$. The result is
\begin{equation}
t_\mathrm{sg} = \frac{1}{\sqrt{2\pi G\rho_0}} = 0.79~\mathrm{Myr} \left(\frac{n_0}{1000~\mathrm{cm}^{-3}}\right)^{-1/2}.
\label{tsg}
\end{equation}
This justifies our choice of measuring post-shock properties at $t=0.2$~Myr.

As explained in \hyperlink{CO14}{CO14}, there are two different length scales (and corresponding characteristic masses) for spherical cores in the post-shock region  at a given ambient density $\rho$: one that is supported by thermal pressure (a critical Bonnor-Ebert sphere)
\begin{align}
R_\mathrm{th,sph} &= 2.3 \frac{c_s}{\sqrt{4\pi G \rho}},\label{Rthsph}\\
M_\mathrm{th,sph} &= 4.18\frac{{c_s}^3}{\sqrt{4\pi G^3 \rho}}, \label{Mthsph}
\end{align}
and one that is supported by magnetic stresses (defined from $4R/3 = L_\mathrm{mag,crit}$)
\begin{align}
R_\mathrm{mag,sph} &= \frac{3}{8\pi\sqrt{G}}\frac{B}{\rho},\label{Rmagsph}\\
M_\mathrm{mag,sph} &= \frac{9}{128\pi^2 G^{3/2}}\frac{B^3}{\rho^2} \label{Mmagsph}
\end{align}
(see Equations~(11)-(12) and (14)-(15) in \hyperlink{CO14}{CO14}).
Figure~\ref{HpsRsph} shows the measured post-shock layer thickness in each model, 
compared with these two possible mass-gathering scales in the post-shock environment, $R_\mathrm{th,sph}$
and $R_\mathrm{mag,sph}$. 
It is obvious from Figure~\ref{HpsRsph} that $R_\mathrm{mag,sph}$ is much larger than the post-shock thickness during the entire core-building phase, and thus magnetically supercritical cores cannot form spherically symmetrically within the post-shock layer. Quantitatively,
since the post-shock layer thickness is $H_\mathrm{ps} = \Sigma_\mathrm{ps}/\left(2 \overline{\rho}_\mathrm{ps}\right)$, we have
\begin{align}
\frac{R_\mathrm{mag,sph}}{H_\mathrm{ps}} &= \frac{3}{4\pi\sqrt{G}}\frac{B_\mathrm{ps}}{\Sigma_\mathrm{ps}} \approx \frac{3}{4\pi\sqrt{G}}\frac{\sqrt{8\pi\rho_0 {v_0}^2} }{2\rho_0 v_0 t}\notag\\
& = \frac{3}{\sqrt{8\pi G \rho_0}t} = 1.2~\left(\frac{n_0}{1000~\mathrm{cm}^{-3}}\right)^{-1/2}\left(\frac{t}{\mathrm{Myr}}\right)^{-1}.
\label{LdHps}
\end{align}
Since the core formation timescales in our models all satisfy $t \lesssim 1$~Myr, Equation~(\ref{LdHps}) suggests $R_\mathrm{mag,sph} > H_\mathrm{ps}$ when cores formed. This means that gravity-induced mass collection in the direction perpendicular to the shocked layer is prevented by magnetic forces, and in-plane mass collection is required for core formation in post-shock regions. 

\subsection{Structure Formation}
\label{sec::psFila}

\begin{figure}
\hspace{-.3in}
\includegraphics[scale=0.13]{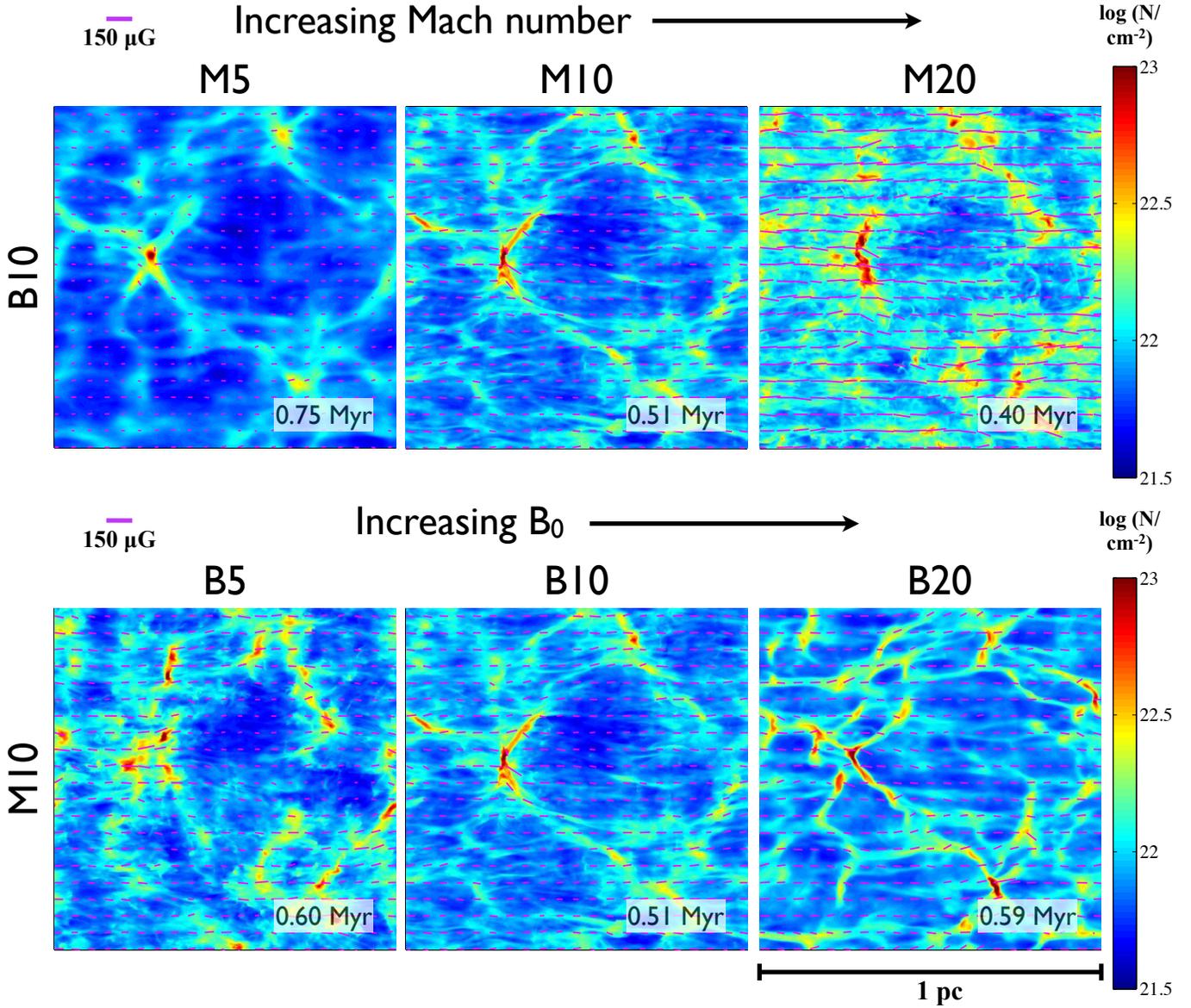}
\caption{An example from one of our simulation runs showing the structure formed in the post-shock layer (in column density; \textit{color map}) for models with different inflow Mach numbers and background magnetic fields. Magnetic field directions in the post-shock layer are also shown ({\it pink segments}). }
\label{spectrumMB}
\end{figure}

\begin{figure}
\begin{center}
\hspace{-.3in}
\includegraphics[width=\textwidth]{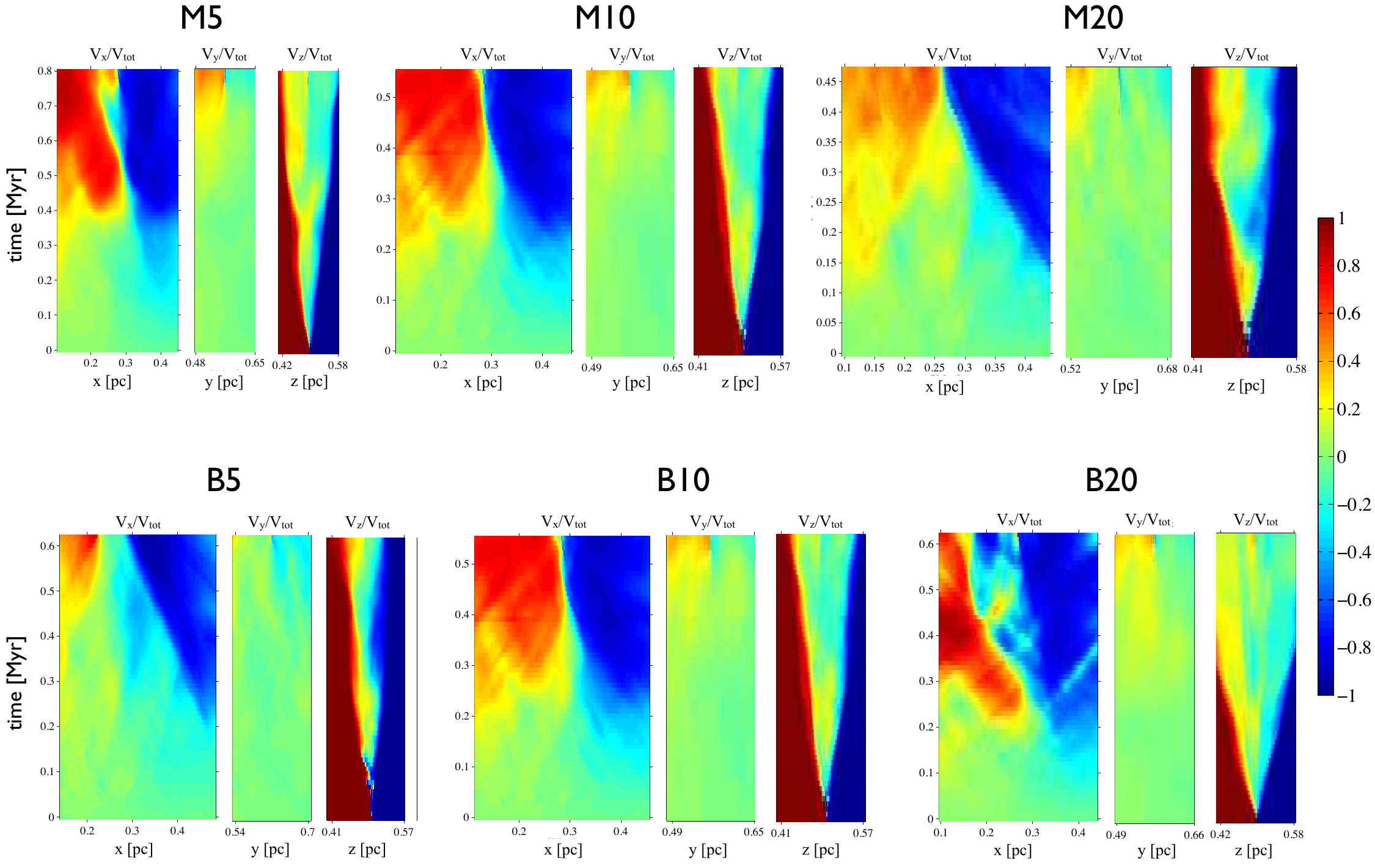}
\caption{The space-time diagrams of $v_x$, $v_y$, $v_z$ around the most evolved core in each model, normalized by the total velocity $v_\mathrm{tot} = \left( {v_x}^2 + {v_y}^2 + {v_z}^2\right)^{1/2}$ at each zone. In all models, $v_z$ dominates in the beginning of the simulation because of the convergent flow setup, but $v_x$ (along the magnetic field lines) soon becomes the strongest component around the forming core. }
\label{spacetime}
\end{center}
\end{figure}

Figure~\ref{spectrumMB} shows examples of structures formed within the post-shock layers, at the time that the most evolved core collapses ($t_\mathrm{coll}$; see Section~\ref{sec::cores}). We have selected models with identical initial turbulence realization, which is responsible for seeding the structures that subsequently grow. Filamentary structures are obviously seen in all models with width $\sim 0.05$~pc, similar to those found in observations \citep[see review in][]{2013PPVI...Andre}. Also, note that the filaments are not necessary perpendicular to the magnetic field, because the locations of nulls in the velocity field are independent of each other on each magnetic field line. 

In addition, we see networks of small sub-filaments or striations parallel to the magnetic field in some models. Similar features have been observed in multiple molecular clouds \citep{2008ApJ...680..428G, 2011ApJ...734...63S, 2012A&A...543L...3H, 2013PPVI...Andre}, and are consistent with the theoretical expectation of anisotropy of magnetized turbulence \citep{1995ApJ...438..763G}. 
Quantitatively, computational studies suggest $\beta\lesssim 0.2$ is required to have significant anisotropy at Mach number $= 5$ \citep{2003ApJ...590..858V,2008ApJ...680..420H}, and the critical value of $\beta$ may become smaller for higher Mach numbers \citep{2008ApJ...680..420H}.
This roughly agrees with our results in Figure~\ref{spectrumMB}: striations parallel to the magnetic field direction (not necessarily perpendicular to the main filaments) are evident in models with low Mach numbers or strong magnetic fields (M5, M10/B10, B20). Otherwise, the high velocity turbulence (M20) or the weak magnetization (B5, see Table~\ref{simresults}) may have destroyed the anisotropy.

Similar to \hyperlink{CO14}{CO14}, we use space-time diagrams of different velocity components to demonstrate the anisotropic process of core formation (Figure~\ref{spacetime}). We consider the region with size $L_x\times L_y \times L_z = L_\mathrm{mag,crit}\times 2 R_\mathrm{th,sph} \times 2 R_\mathrm{th,sph}$ centered around the most-evolved core at $t_\mathrm{coll}$ of each model, and plot the averaged $v_x$, $v_y$, $v_z$ along $x$-, $y$-, $z$-directions in the unit of the total velocity $v_\mathrm{tot}$. Anisotropic gas flows along the $x$-direction are obvious in all models, and appear much earlier than the core collapse (when all three velocity components show convergent flow). Note that, from Figure~\ref{spacetime} we can see that Model B5 has less prominent convergent flow along the $x$-direction than the other models, indicating that anisotropy is not as strong in this model (see Section~\ref{sec::cores}).

\renewcommand{\arraystretch}{1.1}
\begin{table}[t]
\begin{center}
\vspace{-.2in}
  \begin{threeparttable}
\caption{Results from filaments measured at $t=t_\mathrm{coll}$, averaged over all 6 runs for each parameter set.}
\label{FilaSum}
\vspace{.1in}
\begin{tabular}{ l || c c c c c | c c c }
  \hline
  Model & $t_\mathrm{coll}$\tablenotemark{\S} & FFE$_{1.0}$\tablenotemark{\dagger} & FFE$_{1.5}$\tablenotemark{\dagger} & $A_\mathrm{fila, 1.0}$ & $A_\mathrm{fila, 1.5}$ & $L_\mathrm{mag,crit}$ & $L_\mathrm{acc}$\tablenotemark{\ddagger} & $\lambda_m$\tablenotemark{\ddagger} \\ 
    & (Myr) & &  & (pc$^2$) & (pc$^2$) & (pc) & (pc) & (pc) \\
  \hline
  M5B10  & 0.83 & 0.65 & 0.31 & 0.34 & 0.11 & 0.19 & 0.51 & 0.39 \\
  M10B10  & 0.53 & 0.57 & 0.27 & 0.34 & 0.11 & 0.18 & 0.49 & 0.28 \\
  M20B10 & 0.43 & 0.59 & 0.31 & 0.35 & 0.13 & 0.19 & 0.51 & 0.20 \\
  \hline
  M10B5 & 0.58 & 0.61 & 0.33 & 0.34 & 0.13 & 0.10 & 0.27 & 0.28 \\
  M10B10 & 0.53 & 0.57 & 0.27 & 0.34 & 0.11 & 0.18 & 0.49 & 0.28 \\
  M10B20 & 0.63 & 0.53 & 0.36 & 0.30 & 0.15 &  0.37 & 1.00 & 0.28\\
  \hline 
\end{tabular}
    \begin{tablenotes}
      \footnotesize
      \item $^\S$Collapse is defined as the time when $n_\mathrm{max} =10^7$~cm$^{-3}$ in each simulation.
      \item $^\dagger$FFE (filament formation efficiency) is the ratio of the total mass in filamentary structures to the total mass in the shocked layer at $t_\mathrm{coll}$, as defined in Equation~(\ref{FFE}).
      \item $^\ddagger$See Section~\ref{sec::cores}.
    \end{tablenotes}
  \end{threeparttable}
\end{center}
\end{table}

\begin{figure}[t]
\begin{center}
\hspace{-.3in}
\includegraphics[width=\textwidth]{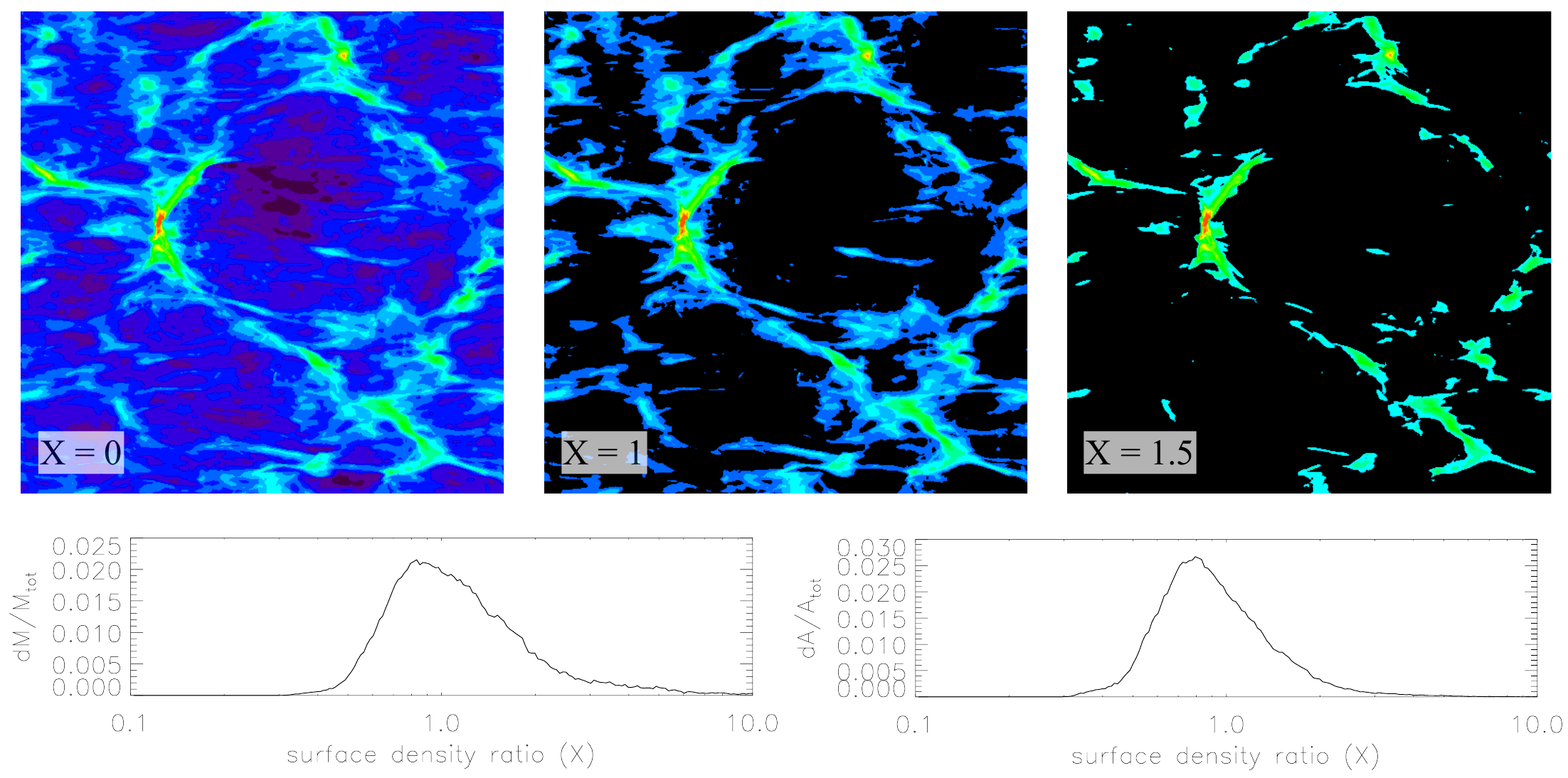}
\caption{Comparison between filamentary structures above different cut-off $X$ values in the criterion $\Sigma > X \cdot \overline{\Sigma}_\mathrm{ps}$ (\textit{top}), and the fraction of filament mass (\textit{bottom left}) and the fraction of filament area (\textit{bottom right}) as functions of $X$, from model M10B10.}
\label{FilaBD}
\end{center}
\end{figure}

Quantitatively, if we define overdense (filamentary) structures as those with surface density contrast higher than a certain value, say, $\Sigma > X \cdot \overline{\Sigma}_\mathrm{ps}$, then we can measure the mean surface density of filaments, $\overline{\Sigma}_\mathrm{fila}$, as the ratio of total mass inside filamentary structures,
\begin{equation}
M_\mathrm{fila} \equiv \int\limits_{\Sigma > X\cdot \overline{\Sigma}_\mathrm{ps}} \Sigma(x,y) ~ dxdy,
\end{equation}
to total area ($A_\mathrm{fila}$) of the same structures.
The filament formation efficiency (FFE) is defined by:
\begin{equation}
\mathrm{FFE} = \frac{M_\mathrm{fila}}{M_\mathrm{ps}} = \frac{M_\mathrm{fila}}{2\rho_0 v_0 t}.
\label{FFE}
\end{equation}
Table~\ref{FilaSum} lists (at time $t = t_\mathrm{coll}$ for each model) the measured FFE and total area of filaments using $X=1.0$ and $X=1.5$, as well as three mass-accreting scales $L_\mathrm{mag,crit}$ (see Equation~(\ref{Lcrit})), $L_\mathrm{acc}$, and $\lambda_m$ (see discussion in Section~\ref{sec::cores}). 
Though the core collapse timescale varies with inflow Mach number, the filament formation efficiency and the total area of filaments do not seem to have strong dependence on either the inflow Mach number or the pre-shock magnetic field. This is in contrast to the core formation efficiency (CFE), which varies with $t_\mathrm{coll}$ (see Table~\ref{CoreSum} and discussion in Section~\ref{sec::cores}).

We have also explored how the FFE varies in time (based on individual models). We find that the FFE it is fairly constant during self-gravitating stages, with $<15\%$ difference from $t=0.5~t_\mathrm{coll}$ to $t_\mathrm{coll}$ in all models. This, in addition to supporting our choice of measuring FFE at $t_\mathrm{coll}$, indicates that the FFE is not strongly sensitive to the exact age of a cloud.

\begin{figure}
\begin{center}
\includegraphics[width=\textwidth]{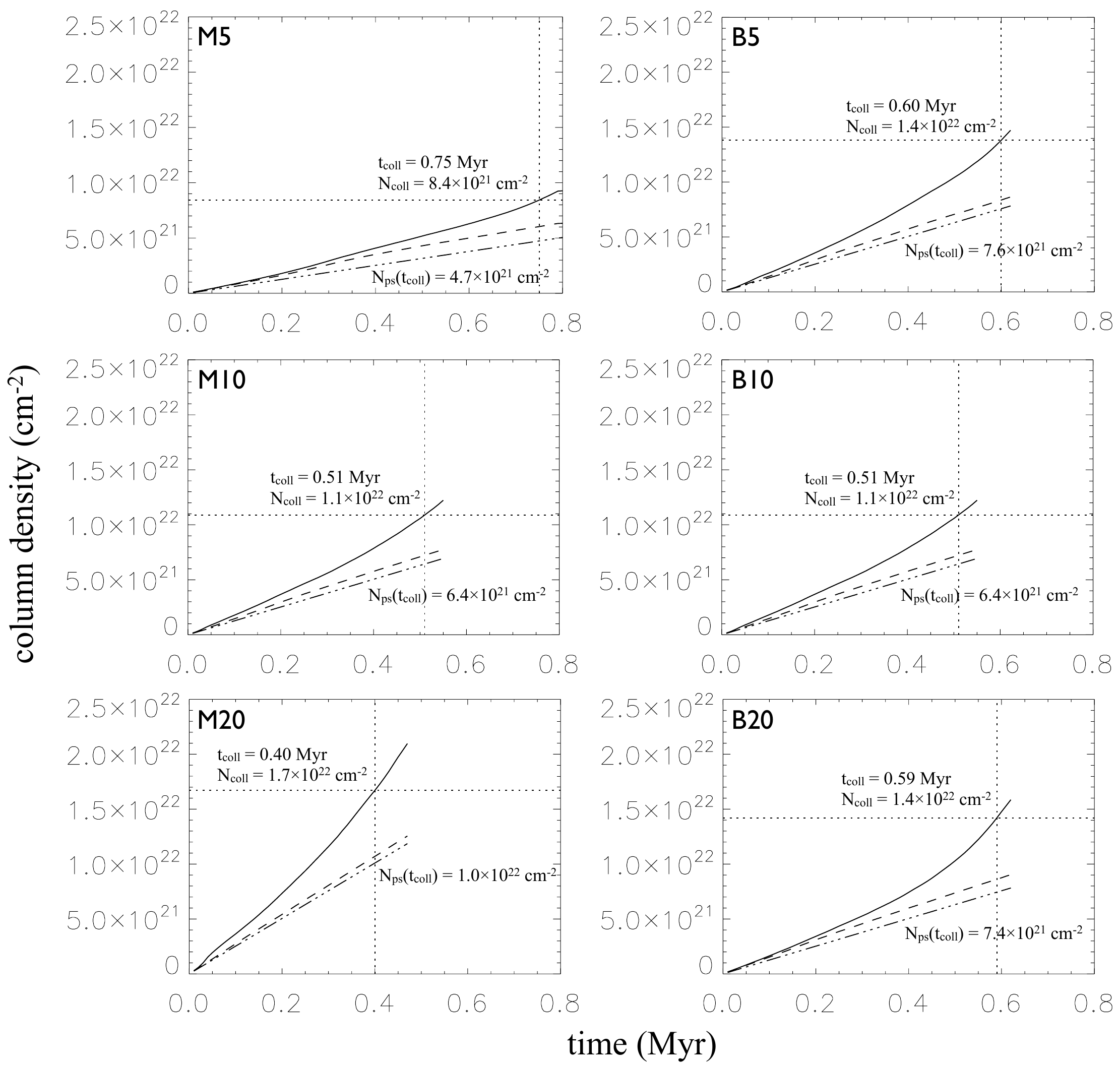}
\caption{An example from one of our simulation runs showing the average column density of the overdense ``filament" structures defined (see text) using $X=1$ (\textit{solid}), post-shock layer (\textit{dashed}), and theoretical value for the post-shock layer (\textit{dash-dotted}) defined as $N_\mathrm{ps}\equiv 2 n_0 v_0 t$. The core collapse time is labeled with dotted lines, with corresponding $N_\mathrm{coll} \equiv N_\mathrm{fila}\left(t_\mathrm{coll}\right)$. }
\label{FilaColDps}
\end{center}
\end{figure}

Note that there is some arbitrariness in the choice of $X$. 
Figure~\ref{FilaBD} compares the post-shock structures under different cutoff values in surface density, and shows the differential PDFs of filament mass and area as functions of the surface density ratio $X \equiv \Sigma / \overline{\Sigma}_\mathrm{ps}$.
Since there is no ``break" in the differential PDF at any particular value of $X$, there is not an obvious value of $X$ to use as a lower limit for filament gas.
Using $X=1.0$, Figure~\ref{FilaColDps} shows $N_\mathrm{fila}$ for each model, as well as the average column density of the post-shock layer, $\overline{N}_\mathrm{ps} = \overline{\Sigma}_\mathrm{ps} / \mu_n$, as functions of time. 
We also measured the filament column density at $t_\mathrm{coll}$; in all models, $N_\mathrm{coll} \equiv N_\mathrm{fila}(t_\mathrm{coll}) \sim 10^{22}$~cm$^{-2}$, comparable to the observed ``critical column density" for filaments with active core formation \citep[see review in][]{2013PPVI...Andre}. In detail, we find that for $X=1$, $N_\mathrm{coll}/\overline{N}_\mathrm{ps} \approx 1.8$ (see Figure~\ref{FilaColDps}).

As we shall show below (see Equation~(\ref{Npstcoll})), the expected post-shock column density at the collapse time is $\overline{N}_\mathrm{ps}\propto \left(n_0 v_0\right)^{1/2}$. When $\Sigma/\overline{\Sigma}_\mathrm{ps} > X=1$ is used to define filaments, $N_\mathrm{coll}/\overline{N}_\mathrm{ps} \approx 1.8$ for all models (see Figure~\ref{FilaColDps}), implying the same dependence of filament column density on $v_0$ as mean post-shock column density, $N_\mathrm{coll}\propto {v_0}^{1/2}$ (see models with different inflow Mach number in the left panel of Figure~\ref{FilaColDps}).

\section{Statistical Core Properties}
\label{sec::cores}

\begin{figure}[t]
\begin{center}
\includegraphics[scale=0.65]{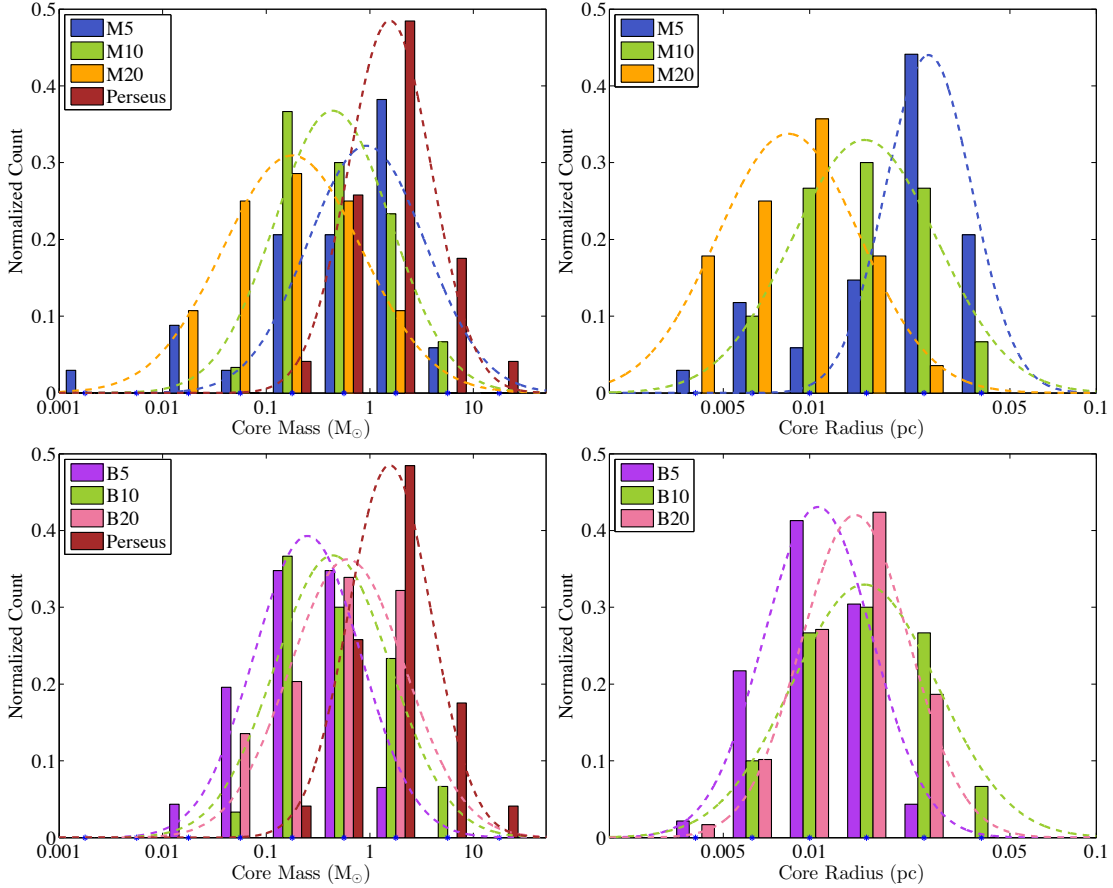}
\caption{Statistical distribution of core mass (\textit{left panel}) and size (\textit{right panel}) for models with different inflow Mach numbers (\textit{top row}) and cloud magnetic fields (\textit{bottom row}). The bin sizes are $10^{0.5}$~M$_\odot$ and $10^{0.2}$~pc, respectively.}
\label{coreMassSizeDist}
\end{center}
\end{figure}

\begin{figure}[t]
\begin{center}
\includegraphics[scale=0.65]{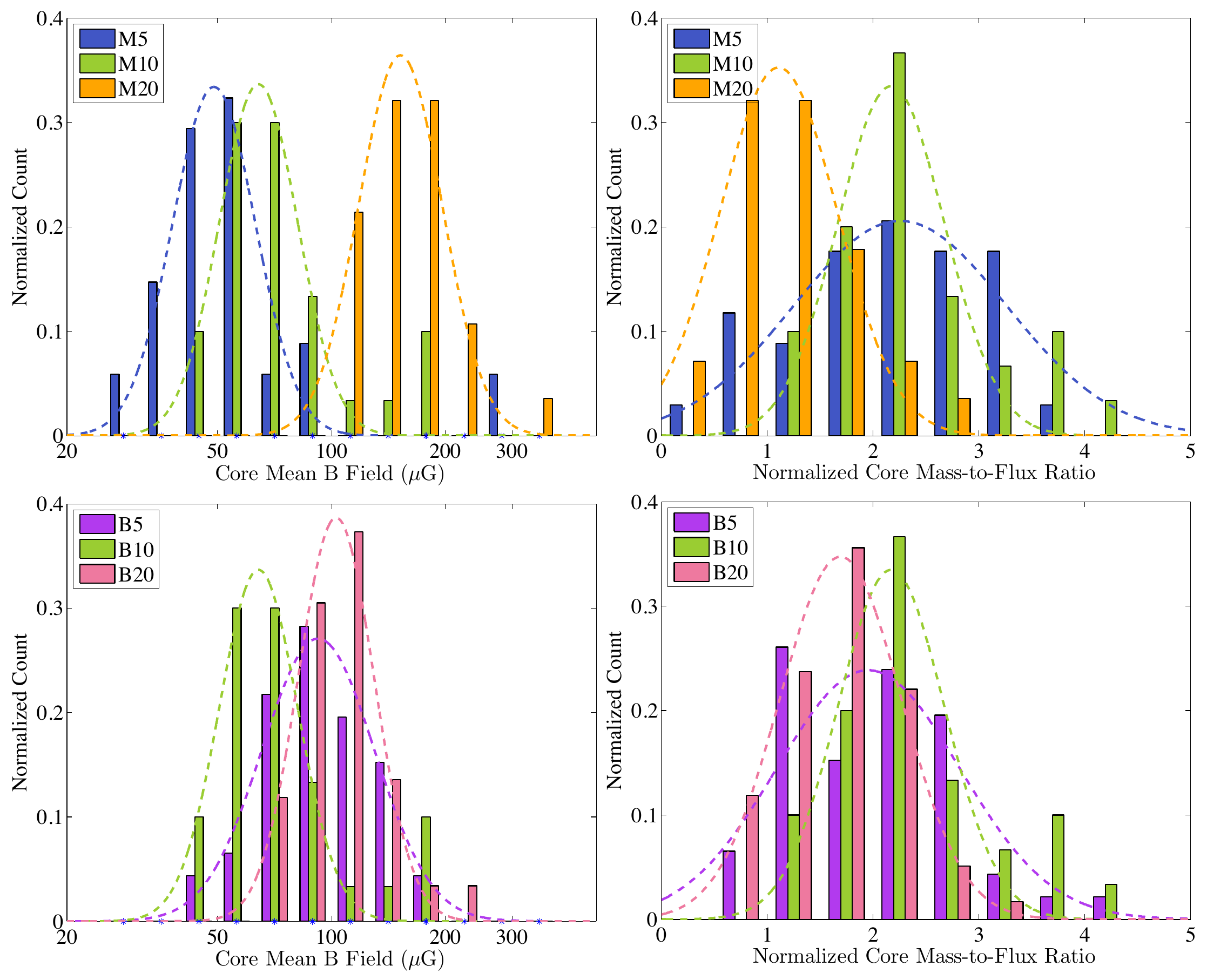}
\caption{Statistical distribution of core mean magnetic field (\textit{left panel}) and mass-to-flux ratio (\textit{right panel}) for models with different inflow Mach numbers (\textit{top row}) and cloud magnetic fields (\textit{bottom row}). The bin sizes are $10^{0.1}~\mu$G and $0.5$, respectively.}
\label{coreMagGammaDist}
\end{center}
\end{figure}

Similar to \hyperlink{CO14}{CO14}, we define the timescale at which $n_\mathrm{max} \geq 10^7$~cm$^{-3}$ as the moment $t_\mathrm{coll}$ when the most evolved core collapses,\footnote{We have tested using $n=10^6$~cm$^{-3}$ and the results are almost the same, since there is little large-scale evolution over that time difference. Based on these tests we found that once a core's maximum density reaches $\sim 10^{6.5}$~cm$^{-3}$ it is definitely collapsing, and therefore we used $n_\mathrm{max} \geq 10^7~\mathrm{cm}^{-3}$ as the criterion of $t_\mathrm{coll}$ in the study.}
then identify cores formed at this time and investigate their physical properties (see Section~\ref{sec::methods}). 
We note that the cores identified in this way correspond to what are termed ``$t_1$ cores" in \cite{2015ApJ...806...31G}. 
Figures~\ref{coreMassSizeDist} and \ref{coreMagGammaDist} show the statistical distributions of core mass, size, mean magnetic field, and mass-to-flux ratio measured from our simulations, normalized by total number of cores identified for each parameter set. The normalized mass-to-magnetic flux ratio is defined as
\begin{equation}
\Gamma \equiv \frac{M}{\Phi_B}\cdot {2\pi\sqrt{G}}.
\end{equation} 
Cores with $\Gamma > 1$ are magnetically supercritical, and have self-gravity strong enough to overcome the magnetic support and collapse.

\begin{figure}[t]
\begin{center}
\includegraphics[scale=0.5]{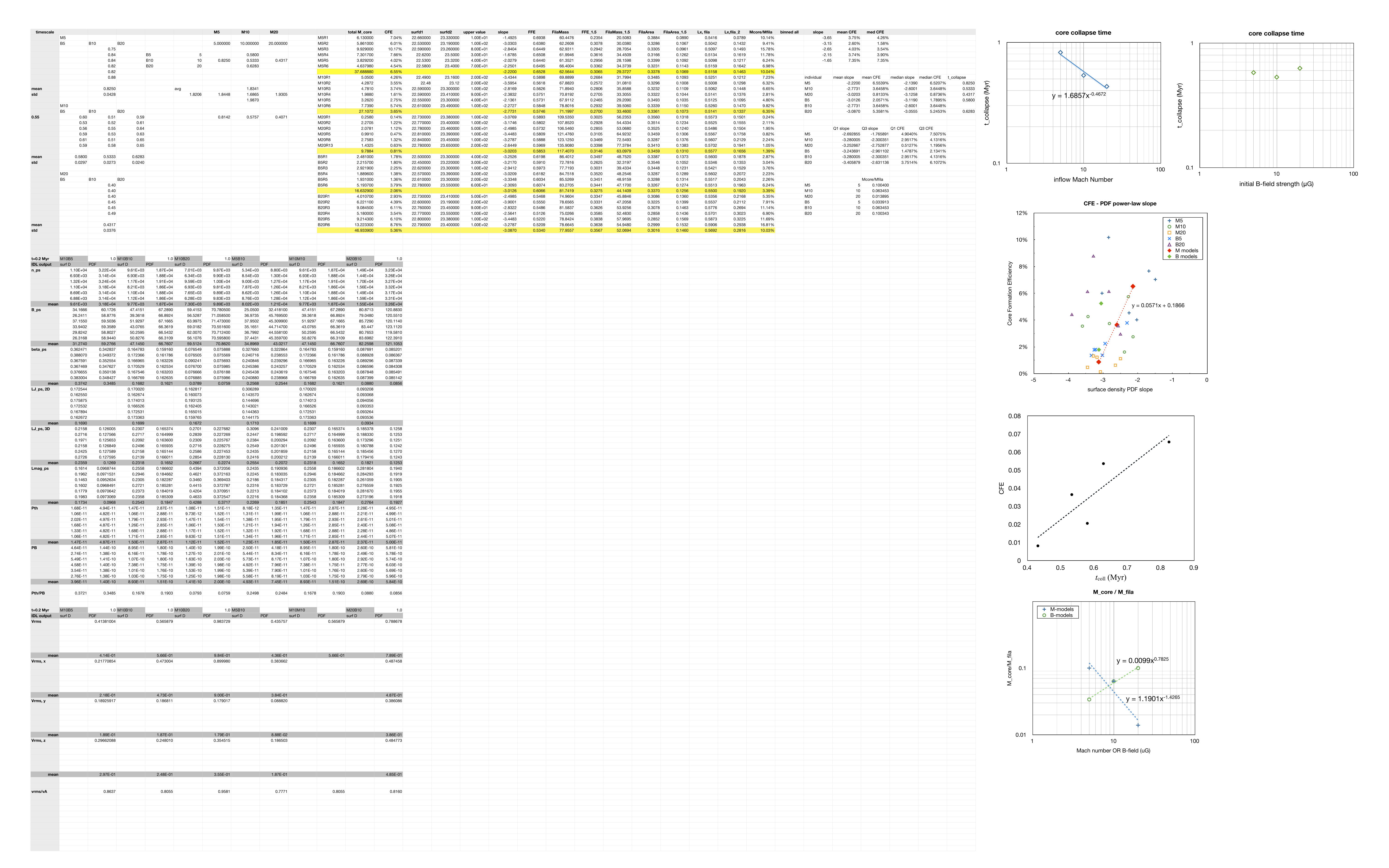}
\caption{The core formation efficiency (CFE) vs. core collapse timescale. Each point represents one model parameter set (M5, M10B10, etc.).}
\label{CFEtcoll}
\end{center}
\end{figure}

Cores identified in our simulations have masses $M_\mathrm{core}\sim 0.002-10$~M$_\odot$, sizes $R_\mathrm{core}\sim 0.004-0.05$~pc, and normalized mass-to-flux ratio $\Gamma\sim 0.4-4.5$, consistent with observations \citep[e.g.][]{2008ApJ...680..457T,2010ApJ...710.1247S,2013MNRAS.432.1424K}. We also included the normalized mass distribution of starless cores in the Perseus molecular cloud \citep[adopted from][]{2010ApJ...710.1247S} in Figure~\ref{coreMassSizeDist} as a comparison (see Section~\ref{sec::obs} for more discussion). The median values of core properties are summarized in Table~\ref{CoreSum}, as well as the averaged core formation efficiency (CFE) and core collapse time $t_\mathrm{coll}$. 
In Figure~\ref{CFEtcoll} we show that the CFE is positively related to the core collapse time, $t_\mathrm{coll}$. This is because more structures in the post-shock region have become nonlinear at later time.
In fact, we found that for individual models, the CFE can increase by an order of magnitude from $t = 0.8~t_\mathrm{coll}$ to $t_\mathrm{coll}$.

Note that though the mean core density, $\overline{n}_\mathrm{core}$, is $\sim 10$~times larger than the ambient density in the post-shock layer, the magnetic field within cores ($\overline{B}_\mathrm{core}$) is not significantly different from the post-shock region (see $\overline{n}_\mathrm{ps}$ and $\overline{B}_\mathrm{ps}$ in Table~\ref{simresults}). This is additional evidence of anisotropic core formation: cores gather material along the magnetic field and become more massive without significantly compressing the field and enhancing the magnetic support. 

\renewcommand{\arraystretch}{1.1}
\begin{table}[t]
\begin{center}
  \begin{threeparttable}
\caption{Results from identified cores measured at $t=t_\mathrm{coll}$.\tablenotemark{\dagger}}
\label{CoreSum}
\vspace{.1in}
\begin{tabular}{ l || c c c | c c c c c c }
  \hline
  Model &  \# Cores & CFE\tablenotemark{\P}& $t_\mathrm{coll}$\tablenotemark{\S} & $\overline{n}_\mathrm{core}$ & $R_\mathrm{core}$\tablenotemark{\ddagger} & $M_\mathrm{core}$ & $\overline{B}_\mathrm{core}$ & $\Gamma_\mathrm{core}$ & $M_\mathrm{core}/M_\mathrm{BE}$ \\ 
   & Identified\tablenotemark{\star} & (\%) & (Myr) & ($10^5$~cm$^{-3}$) & (pc) & (M$_\odot$) & ($\mu$G) &  & \\
  \hline
  M5B10 & 34 & 6.55 & 0.83 & 2.7 & 0.022 & 0.81 &  49 & 2.3 & 2.13\\
  M10B10 & 30 & 3.65 & 0.53 & 4.9 & 0.014 & 0.45 & 69 & 2.1 & 2.46\\
  M20B10 & 28 & 0.81 & 0.43 & 11 & 0.009 & 0.23 & 156 & 1.2 & 1.17\\
  \hline
  M10B5 & 46 & 1.18 & 0.58 & 7.7 & 0.011 & 0.25 & 89 & 2.1 & 0.95\\
  M10B10 & 30 & 3.65 & 0.53 & 4.9 & 0.014 & 0.45 & 69 & 2.1 & 2.46\\
  M10B20 & 59 & 3.90 & 0.63 & 6.9 & 0.015 & 0.55 & 103 &1.7 & 2.66\\
  \hline 
\end{tabular}
    \begin{tablenotes}
      \footnotesize
      \item $^\dagger$Columns (2)$-$(4) are averages over 6 simulation runs for each parameter set. Columns (5)$-$(10) are median values over all cores for each parameter set (6 simulation runs).
      \item $^\star$We only consider gravitationally bound cores with $E_\mathrm{grav} + E_\mathrm{thermal} + E_\mathrm{B} < 0$.
      \item $^\P$CFE is the ratio of the total mass in cores to the total mass in the shocked layer at $t_\mathrm{coll}$ (see Equation~(26) in \hyperlink{CO14}{CO14}).
      \item $^\S$Collapse is defined as the time when $n_\mathrm{max} =10^7$~cm$^{-3}$ in each simulation. The $t_\mathrm{coll}$ shown here is the mean value over all 6 runs for each parameter set.
      \item $^\ddagger$$R_\mathrm{core}$ is calculated from the total number of zones $N$ within a core, for an equivalent spherical volume: $R_\mathrm{core} = (3N/(4\pi))^{1/3}\Delta x$, where $\Delta x=1/512$~pc is the grid size.
    \end{tablenotes}
  \end{threeparttable}
\end{center}
\end{table}

In the anisotropic condensation model (Section~\ref{sec::aniCore}), core properties are expected to depend on the inflow Mach number. In particular, Equations~(\ref{Rcrit}) and (\ref{Mcrit}) suggest that $R_\mathrm{core}$ and $M_\mathrm{core}$ should decrease with increasing ${\cal M}$, while varying $B_0$ should not have significant effect on these core properties. Furthermore, the core field is expected to be comparable to the post-shock value given in Equation~(\ref{Bps}), so that it increases with ${\cal M}$ and is insensitive to $B_0$. Our results in Table~\ref{CoreSum} and Figures~\ref{coreMassSizeDist} and \ref{coreMagGammaDist} generally agree with these theoretical predictions. 

\begin{figure}[t]
\begin{center}
\includegraphics[scale=0.55]{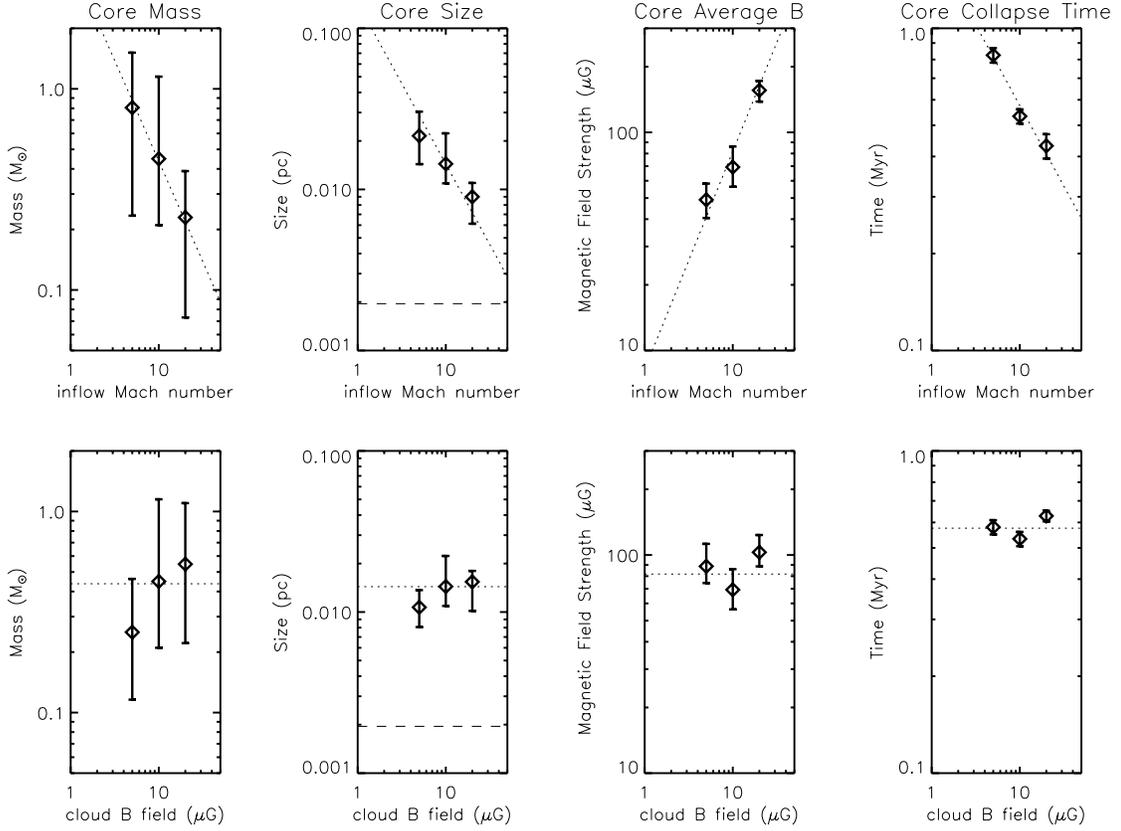}
\caption{Summary of simulated core statistical properties for models with different inflow Mach numbers (\textit{top row}) and cloud magnetic fields (\textit{bottom row}), with theoretical predictions (\textit{dotted lines}). The dashed lines in the core size plots (\textit{second column}) indicate the resolution of our simulations; $\Delta x \approx 0.002$~pc.}
\label{corePropSum}
\end{center}
\end{figure}

Quantitatively, we plot the median values of core mass, size, and mean magnetic field as well as the average core collapse time in Figure~\ref{corePropSum}, as functions of initial Mach number (\textit{top row}) and pre-shock cloud magnetic field (\textit{bottom row}).
We also include theoretical models (\textit{dotted lines}) with $M_\mathrm{core} \propto {\cal M}^{-1}$ (according to Equation~(\ref{Mcrit})), $R_\mathrm{core} \propto {\cal M}^{-1}$ (according to Equation~(\ref{Rcrit})), $B_\mathrm{core} \propto {\cal M}$ (according to Equation~(\ref{Bps})), and $t_\mathrm{coll} \propto {\cal M}^{-1/2}$ (see Equation~(\ref{tcoll}) below). 
For each theoretical comparison, we adopt the predicted scaling and obtain a best-fit coefficient. 
All simulated results fit the theoretical predictions very well, providing quantitative support for the anisotropic core formation model. The fit coefficients we find for radius and mass are $M_\mathrm{core} = 4.4~\mathrm{M_\odot}~{\cal M}^{-1}$ and $R_\mathrm{core} = 0.14~\mathrm{pc}~{\cal M}^{-1}$; these are shown in Figure~\ref{corePropSum}.

The Bonnor-Ebert critical radius and mass for an external pressure $P_\mathrm{ext}$ are given by $R_\mathrm{BE} = 0.485~{c_s}^2 \left( G P_\mathrm{ext}\right)^{-1/2}$ and $M_\mathrm{BE} = 1.2~{c_s}^4 \left( G^3 P_\mathrm{ext}\right)^{-1/2}$. If we take $P_\mathrm{ext} \rightarrow \rho_0 {v_0}^2$ and normalize to $n_0 = 1000$~cm$^{-3}$, $c_s = 0.2$~km/s as in our simulations, the result is
\begin{align}
R_\mathrm{BE,dyn} &= 0.196~\mathrm{pc}\left(\frac{n_0}{1000~\mathrm{cm}^{-3}}\right)^{-1/2}\left(\frac{c_s}{0.2~\mathrm{km/s}}\right){\cal M}^{-1},\notag \\
M_\mathrm{BE,dyn} &= 4.43~\mathrm{M_\odot}\left(\frac{n_0}{1000~\mathrm{cm}^{-3}}\right)^{-1/2}\left(\frac{c_s}{0.2~\mathrm{km/s}}\right)^3{\cal M}^{-1}.
\label{BEdyn}
\end{align}
Comparing to our fitted core radius and mass expressions, we have
\begin{equation}
R_\mathrm{core} = 0.71~R_\mathrm{BE,dyn},\ \ \ M_\mathrm{core} = 0.99~M_\mathrm{BE,dyn}.
\label{coretoBE}
\end{equation}
Therefore, our results suggest that bound core properties are well described by critical Bonnor-Ebert spheres defined by the dynamical pressure of the environment. This supports the key conclusion predicted in our anisotropic core formation model.\footnote{Note that $R_\mathrm{BE,dyn}$ and $M_\mathrm{BE,dyn}$ are respectively factors $0.46$ and $0.43$ smaller than the radius and mass given in Equations~(\ref{Rcrit}) and (\ref{Mcrit}).}

Equations~(\ref{Rcrit}) and (\ref{Mcrit}) were derived assuming that the accumulation length along the magnetic field is $L_\mathrm{mag,crit}$ (Equation~(\ref{Lcrit})). If, however, we instead assume an accumulation length $L_\mathrm{acc}$ and follow the same steps as before, Equations~(\ref{Rcrit}) and (\ref{Mcrit}) would have an additional factor $(L_\mathrm{acc}/L_\mathrm{mag,crit})^{-1}$, i.e.
\begin{equation}
R_\mathrm{core} = 0.43~\mathrm{pc} \left(\frac{n_0}{1000~\mathrm{cm}^{-3}}\right)^{-1/2}\left(\frac{c_s}{0.2~\mathrm{km/s}}\right){\cal M}^{-1} \left(\frac{L_\mathrm{acc}}{L_\mathrm{mag,crit}}\right)^{-1}
\label{RcoreLacc}
\end{equation}
and
\begin{equation}
M_\mathrm{core} = 10.5~\mathrm{M}_\odot \left(\frac{n_0}{1000~\mathrm{cm}^{-3}}\right)^{-1/2}\left(\frac{c_s}{0.2~\mathrm{km/s}}\right)^3{\cal M}^{-1} \left(\frac{L_\mathrm{acc}}{L_\mathrm{mag,crit}}\right)^{-1}.
\label{McoreLacc}
\end{equation}
Comparing to our fits, this implies $L_\mathrm{acc}/L_\mathrm{mag,crit} = 2.4$ or $3.2$ for the mass or radius fit, respectively. 
This suggests that cores actually need to gather material along the magnetic field lines from a length scale $L_\mathrm{acc} > L_\mathrm{mag,crit}$. Since $L_\mathrm{mag,crit}$ represents the critical (minimum) length scale for cores to be magnetically supercritical, our finding of $L_\mathrm{acc} > L_\mathrm{mag,crit}$ is consistent with the anisotropic core formation model. 
Table~\ref{FilaSum} includes the value (in pc) of $L_\mathrm{acc} = 2.7~L_\mathrm{mag,crit}$ in each model that would be required for the median core mass and radius to match Equations~(\ref{McoreLacc}) and (\ref{RcoreLacc}).

We also use the best-fit coefficients found in ${\cal M}$-models (Figure~\ref{corePropSum}, \textit{top row}) to derive the predicted values (\textit{dotted lines}) of core mass, size, magnetic field strength, and collapse time for $B$-models (Figure~\ref{corePropSum}, \textit{bottom row}). Most of the theoretical predictions are in good agreement with the simulation results, except the core mass in the B5 model.
This is because the B5 model has very strong post-shock density compression but only moderate post-shock magnetic field (see Table~\ref{simresults}), and supercritical cores may form isotropically. 
This tendency can also be seen in Figures~\ref{spectrumMB} and \ref{spacetime}, that the structures formed in the B5 model are more randomly distributed compared to other models, the anisotropic gas flow is less prominent, and there is less large-scale structure in the B5 model.

Figure~\ref{corePropSum} shows that the core collapse time follows the relationship $t_\mathrm{coll} \propto {\cal M}^{-1/2}$ very well, as predicted in Equation~(29) of \cite{2011ApJ...729..120G}. The best-fit coefficient gives
\begin{equation}
t_\mathrm{coll}  = 1.82~\mathrm{Myr}~{\cal M}^{-1/2}
\label{tcoll}
\end{equation}
If we compare with Equation~(29) of \cite{2011ApJ...729..120G} (with $n_0 = 1000~\mathrm{cm}^{-3}$ and $c_s = 0.2~\mathrm{km/s}$), this would imply a maximum amplification in the post-shock region of $\ln (\delta\Sigma/\delta\Sigma_0)_\mathrm{max} = 2.29$. The corresponding length scale of the most-amplified mode (see Equation~(30) of \cite{2011ApJ...729..120G}) is then
\begin{align}
\lambda_m & = \left(\frac{2\sqrt{3}\pi}{2.29}\right)^{1/2}\frac{c_s}{\left(G \rho_0\right)^{1/2}}\frac{1}{{\cal M}^{1/2}} \notag \\
& = 0.39~\mathrm{pc} \left(\frac{n_0}{1000~\mathrm{cm}^{-3}}\right)^{-1/2}\left(\frac{v_0}{1~\mathrm{km/s}}\right)^{-1/2}.
\label{lambdam}
\end{align}
In most of our models, $\lambda_m > L_\mathrm{mag,crit}$ (see Table~\ref{FilaSum}), which means the most-amplified mode would be able to form gravitationally bound cores and collapse. 
In fact, the amplification $\ln (\delta\Sigma/\delta\Sigma_0)$ is similar (within $30\%$) for a range of modes with $\lambda$ up to a factor $4$ larger than $\lambda_m$ (see Equation~(26) of \cite{2011ApJ...729..120G}), so it is not surprising that $L_\mathrm{acc}$ differs from $\lambda_m$ (see Table~\ref{FilaSum}).

Using the fitted coefficient of Equation~(\ref{tcoll}) combined with the expectation $t_\mathrm{coll}\propto {n_0}^{-1/2}$, the predicted post-shock surface density at the time of collapse is $\Sigma_\mathrm{ps} \left(t_\mathrm{coll}\right) = 2\rho_0 v_0 t_\mathrm{coll}$, corresponding to column density 
\begin{equation}
N_\mathrm{ps}\left(t_\mathrm{coll}\right) = 5.4\times 10^{21}~\mathrm{cm}^{-2}\left(\frac{n_0}{1000~\mathrm{cm}^{-3}}\right)^{1/2}\left(\frac{v_0}{1~\mathrm{km/s}}\right)^{1/2}. 
\label{Npstcoll}
\end{equation}
This is in good agreement with measured values, as shown in Figure~\ref{FilaColDps}. Considering Equation~(\ref{Npstcoll}) and the fact that $N_\mathrm{fila}\left(t_\mathrm{coll}\right)/\overline{N}_\mathrm{ps} \approx 1.8$ in all models (see Figure~\ref{FilaColDps}), this suggests that the filament column density at the core collapse time may have the same dependence on inflow density and velocity as the post-shock column density, i.e.~$N_\mathrm{fila}\left(t_\mathrm{coll}\right)\propto \left(n_0 v_0\right)^{1/2}$.

\section{Comparison to the Perseus Molecular Cloud}
\label{sec::obs}

\subsection{Cloud Environment}

The dark cloud in Perseus is an active star forming region approximately $250$~pc away, with a total mass of about $10^4~\mathrm{M}_\odot$ over a region about $8\times 25$~pc \citep[see review in][]{2008hsf1.book..308B}. Dense gas tracers and dust emission have revealed filamentary structures and a wealth of dense cores in this region \citep[e.g.][]{2006ApJ...638..293E,2006ApJ...646.1009K}.
In addition, since the Perseus molecular cloud has been observed in $^{12}$CO and $^{13}$CO emission lines \cite[e.g.][]{2006AJ....131.2921R}, the cloud density should be $\gtrsim 10^3$~cm$^{-3}$, similar to the value adopted in our simulations. The Perseus molecular cloud thus represents a good case to compare with our simulation results.

However, the Perseus molecular cloud shows large velocity differences across the region \citep{2008hsf1.book..308B}. The observed CO linewidth is about $5$~km/s over the whole cloud \citep{2006AJ....131.2921R}. Though numerical simulations with rms Mach number ${\cal M} = 6-8$ have shown agreement with observational data on linewidth and cloud structures \citep{1999ApJ...525..318P,2006ApJ...653L.125P}, there is still uncertainty in the actual value of $\sigma_v$ in the Perseus molecular cloud because of the possibility of superposition of multiple clouds \citep{2008hsf1.book..308B}. 

For our comparisons, we adopted the observed properties of starless cores in the Perseus molecular cloud from \cite{2010ApJ...710.1247S}. 
The core mass distribution of Perseus is included in Figure~\ref{coreMassSizeDist} as a comparison to simulations. As discussed in Section~\ref{sec::cores}, the Gaussian-fit peaks of the core mass functions from our simulations shift with the inflow Mach number, or equivalently, the velocity dispersion in the cloud. From Figure~\ref{coreMassSizeDist}, the CMF of Perseus has a peak core mass similar to that of the M5 model, suggesting that Perseus may be a relatively quiescent star-forming environment with converging flow velocities only of order $\sim 1$~km/s.

\subsection{Bonnor-Ebert Mass}

\begin{figure}[t]
\begin{center}
\includegraphics[scale=0.65]{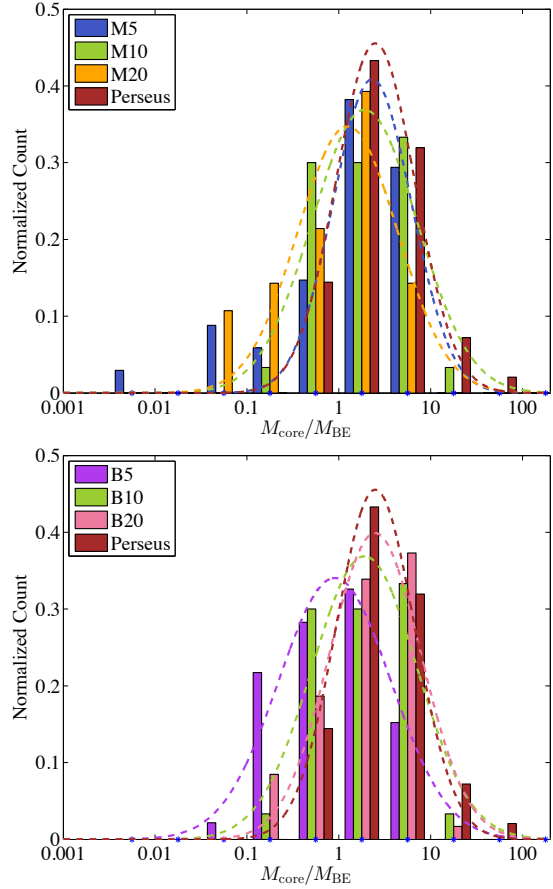}
\caption{Statistical distribution of the ratio between core mass and theoretical Bonnor-Ebert mass, compared with the observed values in Perseus, for cores formed in the M-models (\textit{top}) and B-models (\textit{bottom}).}
\label{BEdist}
\end{center}
\end{figure}

One interesting feature of the Perseus cloud is the existence of ``super-Jeans mass cores" \citep{2010ApJ...718L..32S}. These massive cores have relatively strong self-gravity compared to their internal thermal pressure, but still remain starless. 
An interesting possibility is that these and similar cores may be partially magnetically supported. Our models are useful for addressing this question, because we can measure the fraction of super-Jeans mass cores under different environments in our simulations, and we also can measure magnetic support.

For consistency with theoretical work, we will consider the critical Bonnor-Ebert mass instead of the Jeans mass. We thus convert from the $M/M_J$ ratios in \cite{2010ApJ...718L..32S} to $M/M_\mathrm{BE}$, making use of the core mass and effective radius published in \cite{2010ApJ...710.1247S}, and using Equation~(19) in \cite{2009ApJ...699..230G}:
\begin{align}
M_\mathrm{BE} &= 1.18\frac{{c_s}^4}{\sqrt{G^3 P_\mathrm{edge}}} = 1.85\frac{{c_s}^4}{\sqrt{G^3 P_\mathrm{mean}}}\notag \\
& = 1.85\frac{{c_s}^3}{\sqrt{G^3 \rho_\mathrm{mean}}} = 3.8\frac{{c_s}^3}{G^{3/2}}\frac{R^{3/2}}{M^{1/2}}.
\label{MBE}
\end{align}
For a core at mass $M$, radius $R$, and density $\rho_\mathrm{mean}$ that is pressure confined at its surface, the thermal pressure is insufficient to prevent gravitational collapse if $M > M_\mathrm{BE}$. For each core identified in our simulations, we calculated the value of the critical Bonnor-Ebert mass using the core's mass and radius.

\begin{figure}[t]
\begin{center}
\includegraphics[width=\textwidth]{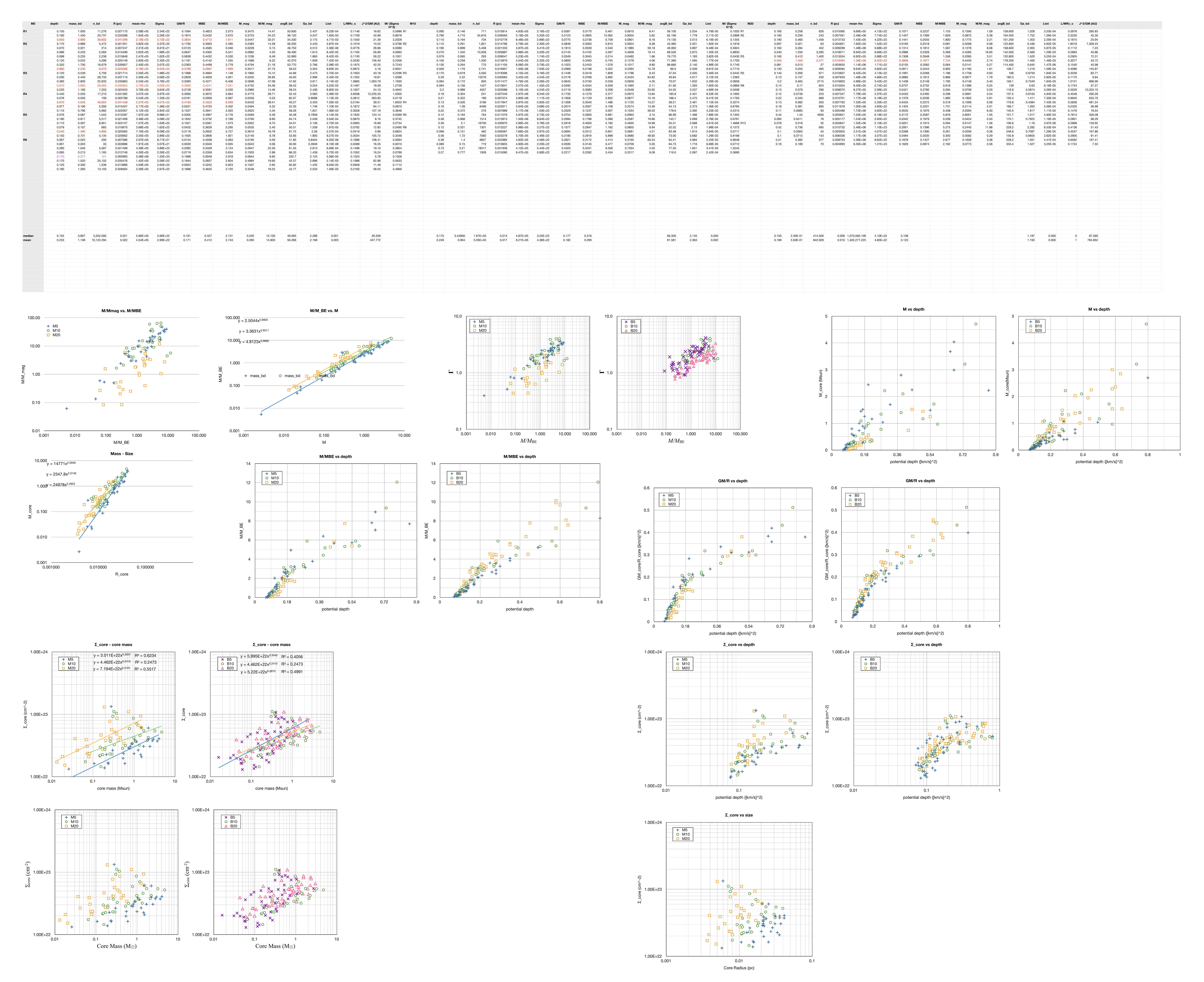}
\caption{Scatter plot of core mass-to-flux ratio vs. $M/M_\mathrm{BE}$ in different models. Each point represents one core formed in the corresponding model.}
\label{GammaMdMBE}
\end{center}
\end{figure}

Figure~\ref{BEdist} shows the statistical distribution of $M_\mathrm{core}/M_\mathrm{BE}$ from both our simulations and Perseus; in addition to the binned counts, we also show best fit lognormal functions for each model. The distributions for all models and for Perseus are similar. Figure~\ref{BEdist} shows that although the median core mass is close to $M_\mathrm{BE}$, the majority of our gravitationally-bound cores have $M_\mathrm{core}/M_\mathrm{BE} > 1$, na\"{i}vely consistent with the fact that these cores are magnetized. 
However, these super-BE mass cores do not in fact seem to be supported primarily by the magnetic field. Figure~\ref{GammaMdMBE} shows the mass-to-flux ratio $\Gamma$ versus $M/M_\mathrm{BE}$ for all cores from our simulations. Evidently, most cores with high $M/M_\mathrm{BE}$ ($\gtrsim 3$) are also strongly magnetically supercritical ($\Gamma \gtrsim 2$). This suggests that the super-BE mass cores observed in Perseus may be strongly self-gravitating and on their way to collapse, rather than being magnetically supported. In fact, in our model, cores with $M_\mathrm{core}/M_\mathrm{BE} \gtrsim 7$ all have $n_\mathrm{max} \gtrsim 10^7~\mathrm{cm}^{-3}$, which means they are the most-evolved collapsing cores in individual simulation runs.

\subsection{Mass-radius Relation}
\label{sec::inplane}

\begin{figure}[t]
\begin{center}
\hspace{-.5in}
\includegraphics[scale=0.5]{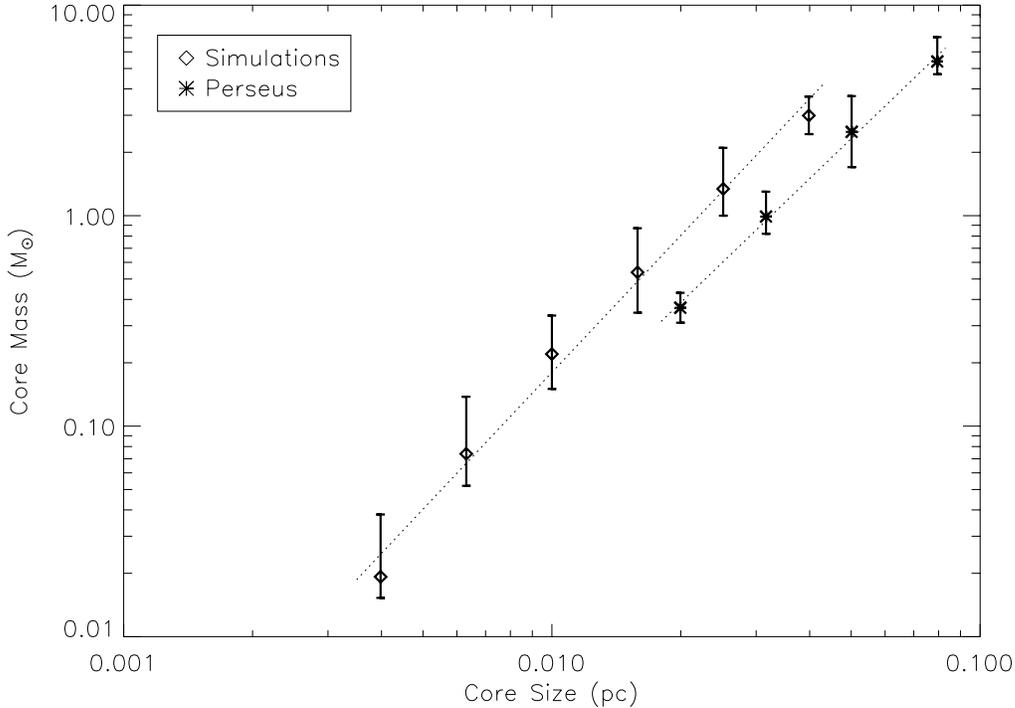}
\caption{The mass-radius relationship measured from our simulations (\textit{diamonds}) compared with the observation results from Perseus (\textit{asterisks}), using the median values of the binned counts. For both the simulations and observations, the vertical bars represent the $\pm 25\%$ values in each bin. The best-fit power laws (\textit{dotted lines}) are $M\propto R^{1.96}$ for Perseus, and $M\propto R^{2.16}$ for our simulations.}
\label{MvsR}
\end{center}
\end{figure}

\begin{figure}[t]
\begin{center}
\includegraphics[width=\textwidth]{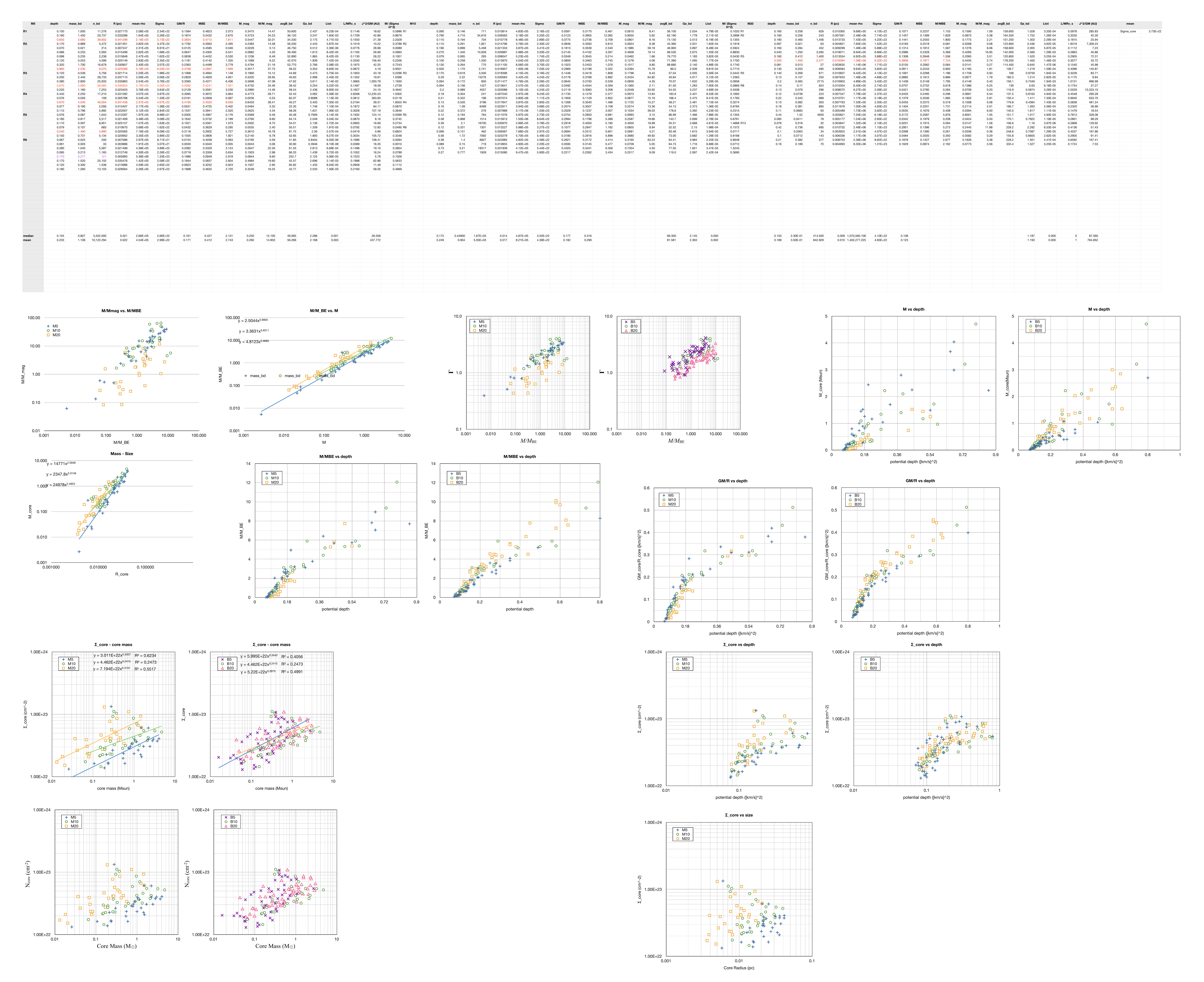}
\caption{Scatter plot of $N_\mathrm{core}$ vs. $M_\mathrm{core}$ in different models. Each point represents one core formed in the corresponding model.}
\label{SigmaMass}
\end{center}
\end{figure}

\begin{figure}
\begin{center}
\includegraphics[width=\textwidth]{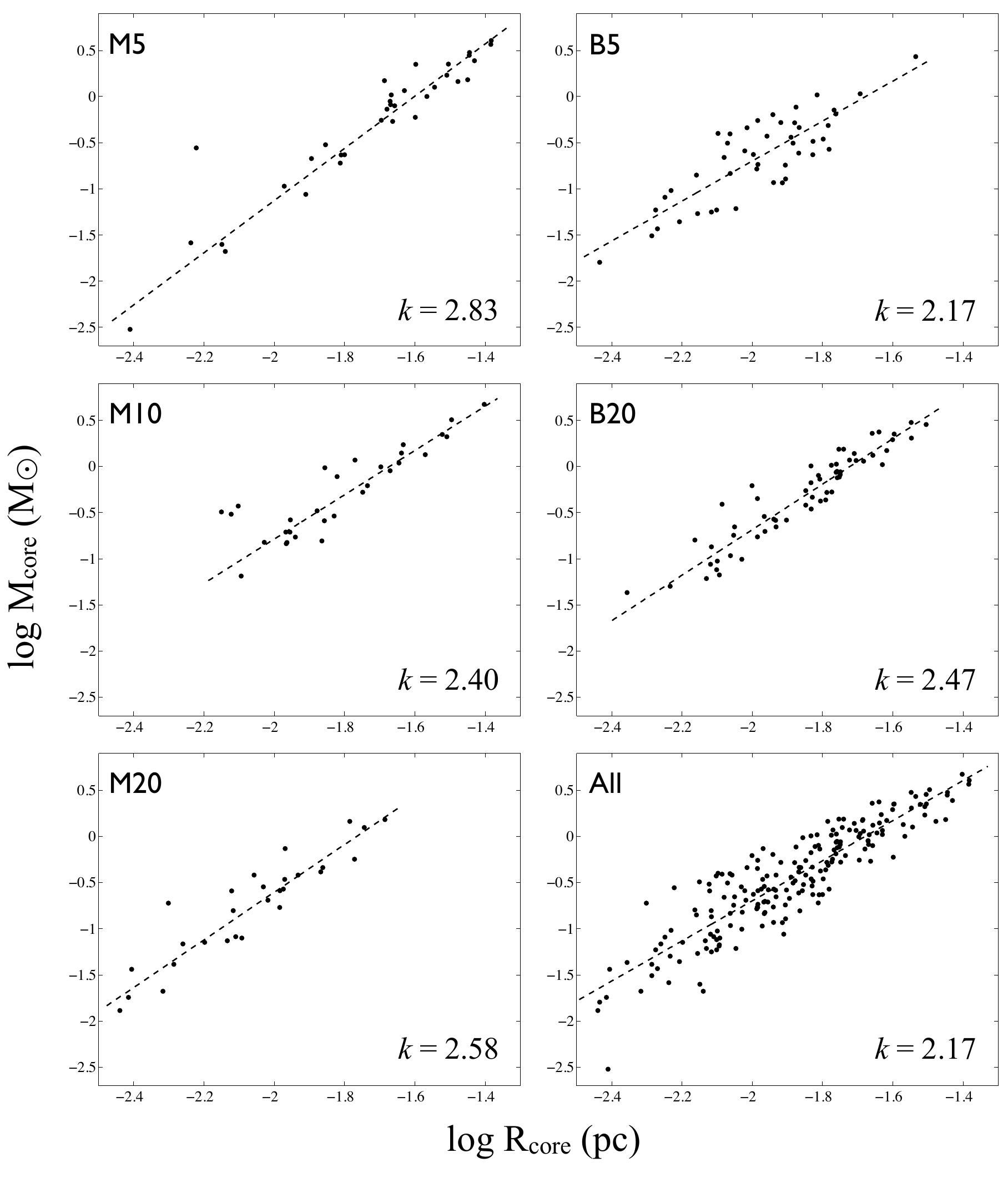}
\caption{Scatter plot of $M_\mathrm{core}$ vs. $R_\mathrm{core}$ in different models (each point represents one core formed in corresponding model), as well as the fitted power-law relationship $M\propto R^k$ (\textit{dashed lines}) with the $k$ values listed on the bottom right of each panel.}
\label{MRfitComp}
\end{center}
\end{figure}

\renewcommand{\arraystretch}{1.1}
\begin{table}[t]
\begin{center}
  \begin{threeparttable}
\caption{The fitted Mass-size relationship from our simulations, $M_\mathrm{core}/M_\odot = A (R_\mathrm{core}/\mathrm{pc})^k$.}
\label{MRfit}
\vspace{.1in}
\begin{tabular}{ l || c c c }
  \hline
  Model & $A~(\times10^3)$ & $k$ & R-square \\ 
  \hline
  M5 & 33.57 & $2.83 \pm 0.18$ & 0.94 \\
  M10 & 10.28 & $2.40 \pm 0.23$ & 0.82 \\
  M20 & 36.14 & $2.58 \pm 0.21$ & 0.86 \\
  B5 & 4.31 & $2.17 \pm 0.22$ & 0.64 \\
  B20 & 17.66 & $2.47 \pm 0.12$ & 0.88 \\
  All & 4.45 & $2.17 \pm 0.08$ & 0.81 \\
  \hline 
\end{tabular}
  \end{threeparttable}
\end{center}
\end{table}

Several observations have found that there is a power-law relationship between the core mass and its size, $M\propto R^k$, with $k\sim 2.4$ \citep{2013MNRAS.432.1424K}. Figure~\ref{MvsR} is the binned mass-size plot from all identified cores in our simulations, compared to the observed cores found in the Perseus molecular cloud \cite[reported in][]{2010ApJ...710.1247S}.
Similar to the observations, the binned data from our simulations show $k \sim 2$ for the power-law relationship between core mass and radius. At a given radius, our cores have slightly higher mass than those in Perseus.

A relationship $M_\mathrm{core} \propto {R_\mathrm{core}}^2$ would suggest that the core surface density $\Sigma_\mathrm{core} \equiv M_\mathrm{core}/(\pi {R_\mathrm{core}}^2)$ is constant for cores regardless of their masses and sizes. Figure~\ref{SigmaMass} shows the scatter plot of the core column density ($N_\mathrm{core} \equiv \Sigma_\mathrm{core}/\mu_n$) versus core mass for all cores formed in our simulations. Although core mass varies over nearly three orders of magnitude ($\sim 0.01 - 10$~M$_\odot$), $N_\mathrm{core}$ is within a factor of $10$. The mean value is $N_\mathrm{core} = 3.7\times 10^{22}$~cm$^{-2}$. By comparison, we found in Section~\ref{sec::psFila} that the overdense filamentary structures have column density $N_\mathrm{fila} \sim 10^{22}$~cm$^{-2}$ at the time of collapse. Thus, the typical core column density $N_\mathrm{core} \sim 4~N_\mathrm{fila}$. 

However, any $k>2$ value indicates that $N_\mathrm{core}$ increases with $M_\mathrm{core}$ or $R_\mathrm{core}$, and this trend is evident in Figure~\ref{SigmaMass}, for different models. Figure~\ref{MRfitComp} shows the mass-radius relations for individual models in our simulations as well as the fitted $M_\mathrm{core} \propto {R_\mathrm{core}}^k$ power-law (the complete fitting coefficients are listed in Table~\ref{MRfit}). We found that the fitted $k$ values are generally higher than $2$, implying that $N_\mathrm{core}$ is not a constant over cores with different masses and sizes.\footnote{Composite distribution of cores from different models show a smaller value of $k$, and more dispersion, than individual models.} 
This means that it is possible that there is no ``universal" core column density, but simply a weak dependence of $N_\mathrm{core}$ on parameters, which is difficult to identify from the present models. For example, the post-shock column density at the time of core collapse varies as $N_\mathrm{ps}\propto \left(n_0 v_0\right)^{1/2}$ (see Equation~(\ref{Npstcoll})), and filament column densities appear to follow a similar trend. If the mean core column density is also a multiple of this, then it would vary by only a factor two for our models, which all have $n_0=1000$~cm$^{-3}$ and have $v_0$ varying by a factor four. We do indeed find a higher mean $N_\mathrm{core}$ for $v_0 = 4$~km/s ($4.8\times 10^{22}$~cm$^{-2}$) compared to $v_0=1$~km/s ($3.0\times 10^{22}$~cm$^{-2}$). Further investigations, both observational and computational, are needed to reach a clearer conclusion.

\section{Summary}
\label{sec::summary}

In this paper, we extended the investigation of \hyperlink{CO14}{CO14} to further examine the anisotropic core formation model and test the theoretical scalings of core properties over a larger parameter space. We carried out fully three-dimensional ideal MHD simulations with self-gravitating gas, including supersonic convergent flows with local turbulence. Our models allow for varying inflow Mach number and magnetic field strength of the background cloud. 
Our simulation results demonstrate that the ram pressure of the converging flow ($\rho_0 {v_0}^2$) is the dominant factor controlling the physical properties of cores formed in the shocked layer. These core properties are consistent with the predictions of the anisotropic core formation theory. Although the post-shock layer is strongly magnetized in all cases, core properties are insensitive to the pre-shock magnetic field strength. We also compared cores formed in our simulations with those observed in the Perseus molecular cloud, and found very similar core mass distribution, super-Bonnor-Ebert mass ratio, and mass-size relation.

Our main conclusions are as follows: 
\begin{enumerate}

\item Considering typical GMC conditions, spherically symmetric core formation is impossible in the magnetized post-shock region, because the required mass gathering scales are much larger than the thickness of the shocked layer (Table~\ref{simresults} and Figure~\ref{HpsRsph}). Quantitatively, it takes $\gtrsim 1$~Myr for the post-shock layer thickness to be comparable with the magnetic critical length under post-shock conditions (Equation~(\ref{LdHps})), much longer than typical core formation timescale in our simulations.

\item Filamentary structures formed in the post-shock regions are similar to those found in observations, with dense cores embedded within filaments (Figure~\ref{spectrumMB}). We measured the filament formation efficiency (FFE) to be around $50\%$ (dependent on the choice of column density threshold of filament; Table~\ref{FilaSum}), independent of the pre-shock conditions.
We also found that the filament column density at the time when cores start to collapse is proportional to the mean post-shock column density; $N_\mathrm{fila}\left(t_\mathrm{coll}\right) \approx 1.8 \overline{N}_\mathrm{ps}$ (Figure~\ref{FilaColDps}).

\item Our velocity space-time diagrams (Figure~\ref{spacetime}) show clear evidence that the mass-gathering flows that create cores and filaments are highly anisotropic. Until late times, flow along the magnetic field is much stronger than in the two perpendicular directions. However, our simulations also show that the ``seeds" of cores are present even at early times. This suggests that core and filament formation is simultaneous, instead of the commonly-assumed picture that cores form only after filaments do.

\item Magnetically supercritical cores form within the post-shock layers in all of our simulations, with masses $\sim 0.002-10$~M$_\odot$, sizes $\sim 0.004-0.05$~pc, and normalized mass-to-flux ratio $\sim 0.4-4.5$ (Table~\ref{CoreSum}). The core formation timescale is $t_\mathrm{coll} \sim 0.4-0.9$~Myr, and the core formation efficiency is positively-related to the core collapse time (Figure~\ref{CFEtcoll}). 

\item The statistical distributions of core mass, size, mean magnetic field, and mass-to-flux ratio clearly show that median core properties depend on the pre-shock inflow Mach number ${\cal M} = v_0 /c_s$ but not the upstream magnetic field strength $B_0$ (Figures~\ref{coreMassSizeDist} and \ref{coreMagGammaDist}). The theoretical scalings predicted in the anisotropic core formation model are $M_\mathrm{core} \propto {\cal M}^{-1}$, $R_\mathrm{core} \propto {\cal M}^{-1}$, and $B_\mathrm{core} \sim B_\mathrm{ps} \propto {\cal M}$ (Equations~(\ref{Bps})-(\ref{Mcrit})), which agree with our simulation results very well (Figure~\ref{corePropSum}). Furthermore, the core collapse timescale in our MHD simulations generally follow the relationship $t_\mathrm{coll}\propto {\cal M}^{-1/2}$. The $t_\mathrm{coll}$ scaling is consistent with the prediction of \cite{2011ApJ...729..120G} based on hydrodynamic analysis, because the flows in the post-shock layer are primarily parallel to the magnetic field. This also gives the post-shock column density at $t_\mathrm{coll}$ to be $N_\mathrm{ps}\left(t_\mathrm{coll}\right) \propto {\cal M}^{1/2}$ (Equation~(\ref{Npstcoll})).

\item Quantitatively, the median core mass and radius depend on inflow velocity as $M_\mathrm{core} = 0.88~\mathrm{M}_\odot\left({v_0}/\left(\mathrm{km/s}\right)\right)^{-1}$ and $R_\mathrm{core} = 0.028~\mathrm{pc}\left({v_0}/\left(\mathrm{km/s}\right)\right)^{-1}$. This suggests that the core mass and radius will be, respectively, a factor $0.99$ and $0.71$ lower than the Bonnor-Ebert critical mass and radius computed using the sound speed and total dynamical pressure ($\rho_0 {v_0}^2$) in the cloud (Equations~(\ref{BEdyn}) and (\ref{coretoBE})). This result is similar to the scaling for characteristic mass proposed by \cite{1997MNRAS.288..145P}, but our measured coefficient is higher by a factor $\sim 2$.

\item Cores identified in our simulations have physical properties very similar to those observed in Perseus \citep{2010ApJ...710.1247S}. In addition, we found similar statistical distributions of $M_\mathrm{core}/M_\mathrm{BE}$ in simulations and observations (Figure~\ref{BEdist}). We suggest that the ``super-Bonnor-Ebert mass cores" identified in \cite{2010ApJ...718L..32S} are probably not supported by magnetic pressure and will collapse gravitationally, since most cores with high $M_\mathrm{core}/M_\mathrm{BE}$ in our simulations also have high $\Gamma$ values, indicating that these cores are magnetically supercritical (Figure~\ref{GammaMdMBE}).

\item We find (Figure~\ref{MvsR}) a composite mass-radius relation for our simulated prestellar cores comparable to that seen in observations, $M_\mathrm{core} \propto {R_\mathrm{core}}^k$ with $k=2-2.5$ \citep[e.g.][]{2013MNRAS.432.1424K}. Although the observed relation is sometimes interpreted as implying a ``universal" core surface density, our results suggest that there might be a weak dependence of the core surface density $\Sigma_\mathrm{core} \equiv M_\mathrm{core}/\left(\pi {R_\mathrm{core}}^2\right)$ on core mass or radius (Figure~\ref{SigmaMass}). We also find that the exponent $k$ in the mass-size relation $M \propto R^k$ is larger for individual models with consistent shock conditions than the composite from heterogeneous environments (Figure~\ref{MRfitComp}).

\end{enumerate}

To conclude, 
the success of the anisotropic core formation model for explaining idealized converging turbulent magnetized flows is very encouraging, and provides strong motivation for testing these ideas in global MHD simulations of star-forming molecular clouds.
Further investigations considering more extreme conditions of GMCs would also be interesting to examine the properties of core-forming filaments, and potential variations in the core mass-size relationship.

\acknowledgements
This work was supported by NNX10AF60G from NASA ATP, and by grant NNX13AO52H supporting C.-Y. C. under the NASA Earth and Space Science Fellowship Program. We are grateful to Sarah Sadavoy for providing a table of core properties in Perseus,
and to the referee for a detailed and thorough report that helped us to improve the manuscript.

\end{document}